\documentstyle[amssymb,preprint,aps,prb,epsf,epsfig]{revtex} 
 
\tighten 
\ifpreprintsty  
\makeatletter 
\def\@CITEX[#1]#2{\if@filesw\immediate\write\@auxout{\string\citation{#2}}\fi 
\leavevmode [\@cite{\@collapse{#2}}{#1}]} 
\makeatother 
\fi 
\newcommand{\Sq}{\smash{\hat{\vec{S}}}{\vphantom{\hat{S}}\,}^2}  
\begin{document} 
\draft 
 
\title{\centerline{Spin-Waves in itinerant ferromagnets}} 
 
\author{J.~B\"unemann} 
\address{Fachbereich.~Physik, Universit\"at Marburg, D-35032 Marburg, Germany} 
 
\maketitle 
\begin{abstract}%
 
We introduce a novel approach for the investigation of spin-wave excitations in itinerant ferromagnets. Our theory is based on a variational treatment of general multi-band Hubbard models which describe elements and compounds of transition metals. The magnon dispersion is derived approximately as the energy of a variational spin-wave state in the limit of large spatial dimensions. A numerical evaluation of our results is feasible for general multi-band models. As a first application we consider a model with two degenerate orbitals per lattice site. From our results we can conclude that spin-wave excitations in strong itinerant ferromagnets are very similar to those in ferromagnetic spin systems.   
 
\end{abstract} 
 
 
\section{Introduction} 
\label{intro} 
The most prominent examples for metals with a ferromagnetic order are the 
elements of the iron group, namely iron, cobalt and nickel. Although the 
magnetic behaviour of these materials is a well-known phenomenon, there are 
still many open questions in this field (for a general introduction, see 
e.g., Refs. \onlinecite{Herring}, \onlinecite{Fazekas} ). It is generally 
accepted that the basic reason for ferromagnetic order is the interplay 
between the kinetic energy and the Coulomb-interaction of the electrons. 
Nevertheless, it is still a matter of debate which kind of minimal model 
must be used for the description of ferromagnetic materials. Whatever an 
appropriate Hamilton may be, from a theoretical point of view one would 
expect that it will be a hard analytical task to find even an approximate 
solution for such a real many-particle problem. 
 
The simplest theory, which gives an explanation for metallic ferromagnetism 
is the Hartree-Fock-Stoner theory \cite{stoner}. A surprising result of this 
theory is the statement that ferromagnetism occurs in any system provided 
that the product of Coulomb-interaction and the electrons' density of states 
exceeds a certain amount. This statement is the famous Stoner criterion. 
This criterion usually leads to surprisingly small critical values of the 
Coulomb-interaction for which ferromagnetism is predicted to occur. At first 
sight one may consider this as an a posteriori justification of the 
Hartree-Fock theory, which certainly fails for stronger Coulomb 
interactions. However, in some simpler model systems like the one-band 
Hubbard model it is well known that ferromagnetism requires very large 
Coulomb-interactions if it exists at all. This is a definite contradiction 
to Hartree-Fock theory, which indicates that the whole Stoner picture may be 
inadequate for a thorough understanding of itinerant ferromagnetism. Similar 
objections could be raised against spin-density functional theory which, 
like Stoner-theory, is based on an effective one-particle description. 
Despite their conceptual shortcomings, these theories quite successfully 
describe some properties of the iron-group elements, i.e. the magnetic 
moment or the shapes of the multisheet Fermi surfaces \cite{Moruzzi}. 
Therefore, a competitive strong-coupling theory must meet these apparent 
successes of spin-density functional theory before it will be taken 
seriously. 
 
To this end, we recently introduced a variational treatment of multi-band 
Hubbard models with a general class of Gutzwiller wave functions \cite 
{Gutzwiller1963,PRB98}. These models allow the description of real 
materials, for example the elements of the iron group. As a first 
application we studied the ferromagnetic transition in a two-band model. In 
contradiction to the Hartree-Fock theory we found that ferromagnetism 
requires quite large Coulomb interactions. In particular, we demonstrated 
the decisive role of the intra-atomic exchange interaction, which is found 
to be irrelevant in the Hartree-Fock approach. The Gutzwiller theory shows 
that finite values of this exchange interaction are essential for metallic 
ferromagnetism. Based on these results, we suggested that the complex atomic 
Coulomb-interaction has to be taken seriously in theories on itinerant 
ferromagnetism. In particular, it appears to be essential to take into 
account exchange interactions which form local spins according to Hund's 
rule. We are presently calculating physical properties of the iron-group 
elements from our correlated electron approach \cite{WGB}. First results for 
Nickel show that our method is able to resolve all major discrepancies 
between experiment and spin-density functional theory\cite{BGW-Gutzi}. 
 
Experiments not only provide information about ground-state physics, but 
also yield insight into dynamical properties of materials. It is found that 
metallic and insulating ferromagnets behave similar with respect to their 
low-energy spin excitations. In both cases, inelastic neutron-scattering 
experiments show pronounced gapless spin-wave excitations\cite{Mook,Lowde}. 
The understanding of these excitations is very important since they govern 
the magnetic phase transition at finite temperatures\cite{Morija}. 
Theoretical methods which allow the determination of spin-wave dispersions 
in Hubbard models are quite rare. It is obvious that a convincing 
description of spin excitations can only be obtained starting from a 
qualified theory for the ground-state. However, almost all earlier theories 
on spin-wave excitations in ferromagnets are based on effective one-particle 
theories. Some of these theories lead to surprisingly accurate results 
compared to experiments. For example, in Ref. \onlinecite{Cook} a 
random-phase approximation for iron and nickel was introduced. In other 
approaches, the spin-wave dispersion is determined via a mapping of the 
itinerant system to a ferromagnetic Heisenberg model (see, e.g. Ref. %
\onlinecite{Eschrig}). 
 
In this work, we present a theory for spin-wave excitations in itinerant 
ferromagnets which is based on the variational ground states in Ref. %
\onlinecite{PRB98}. Our paper is organized as follows: In Sec. \ref{hamilt} 
we introduce the general class of multi-band Hubbard models and the 
corresponding Gutzwiller wave functions. Our general approach on the 
spin-wave problem is presented in Sec. \ref{spin}. In Sec. \ref{calcu} we 
evaluate the spin-wave dispersion for our general class of multi-band 
Hubbard models. Finally, we apply our results to a model with two degenerate 
bands in Sec. \ref{zweiband}. Short conclusions close our presentation. 
Technical details are deferred to four appendices.

\section{Hamiltonian and variational wave function} 
\label{hamilt} 
\subsection{Multi-band Hamiltonian \label{sec2.1}} 
 
In this paper we consider the following general class of multi-band Hubbard 
models,  
\begin{equation} 
\hat{H}=\sum_{i(\neq )j}\sum_{\sigma ,\sigma ^{\prime }}t_{i,j}^{\sigma 
,\sigma ^{\prime }}\hat{c}_{i;\sigma }^{+}\hat{c}_{j;\sigma ^{\prime }}^{%
\vphantom{+}}+\sum_i\hat{H}_{i;\text{at}}\equiv \hat{H}_1+\hat{H}_{\text{at}%
}\;.  \label{1.1} 
\end{equation} 
Here, $\hat{c}_{i;\sigma }^{+}$ creates an electron with combined spin-orbit 
index~$\sigma =1,\ldots ,2N$ ($N=5$ for 3$d$~electrons) at the lattice site~$%
i$ of a solid. For simplicity, we assume that the orbitals do not belong to 
the same representation of the respective point-symmetry group. For example, 
in cubic symmetry this means that there is only one set of $s,p,e_g$ and $%
t_{2g}$-orbitals. In this case, one-particle-states $|\Phi _0\rangle $, 
which respect the symmetry of the lattice, lead to vanishing non-diagonal 
local hopping-terms, i.e.  
\begin{equation} 
\left\langle \Phi _0\left| \hat{c}_{i,\sigma }^{+}\hat{c}_{i,\sigma ^{\prime 
}}^{}\right| \Phi _0\right\rangle \sim \delta _{\sigma ,\sigma ^{\prime }}\;. 
\label{1.1b} 
\end{equation} 
This relation simplifies the calculations in this paper, but there is no 
fundamental obstacle to extend our method to a more general case. 
 
We further assume that the atomic Hamiltonian  
\begin{equation} 
\hat{H}_{i;\text{at}}=\sum_\sigma \epsilon _\sigma \hat{n}_\sigma 
+\sum_{\sigma _1,\sigma _2,\sigma _3,\sigma _4}{\mathcal{U}}^{\sigma 
_1,\sigma _2;\sigma _3,\sigma _4}\hat{c}_{\sigma _1}^{+}\hat{c}_{\sigma 
_2}^{+}\hat{c}_{\sigma _3}^{\vphantom{+}}\hat{c}_{\sigma _4}^{\vphantom{+}%
}\;.  \label{1.2} 
\end{equation} 
is site-independent and readily diagonalized,  
\begin{mathletters} 
\begin{eqnarray} 
\hat{H}_{\text{at}} &=&\sum_\Gamma E_\Gamma \hat{m}_\Gamma \;,  \label{1.3} 
\\ 
\hat{m}_\Gamma &=&\left| \Gamma \right\rangle \left\langle \Gamma \right| \;. 
\end{eqnarray} 
Here, we introduced the eigenvalues $E_\Gamma $ and the eigenstates $\left| 
\Gamma \right\rangle $ of $\hat{H}_{\text{at}}$. The diagonalization of $%
\hat{H}_{\text{at}}$ is a standard exercise (see e.g., Ref. %
\onlinecite{Sugano}). Knowledge of the states $\left| \Gamma \right\rangle $ 
means that we found their expansion  
\end{mathletters} 
\begin{equation} 
|\Gamma \rangle =\sum_IT_{I,\Gamma }|I\rangle \;,  \label{1.4} 
\end{equation} 
in the basis of the configuration states $\left| I\right\rangle $. In these 
states, a definite set of spin-orbit states $\sigma $ is occupied  
\begin{equation} 
\left| I\right\rangle =\left| \sigma _1,\sigma _2,...\right\rangle =\hat{c}%
_{\sigma _1}^{+}\hat{c}_{\sigma _2}^{+}\cdot \cdot \cdot |\text{vacuum}%
\rangle \;\;\;\;\;(\sigma _1<\sigma _2<...). 
\end{equation} 
For details about the notation, see Ref. \onlinecite{PRB98}, Sec. II. 
 
\subsection{Gutzwiller-wave-function and diagrammatic evaluation} 
 
In Ref. \onlinecite{PRB98} we proposed the following wave-function for a 
variational examination of the Hamiltonian (\ref{1.1}):  
\begin{equation} 
|\Psi _{\text{G}}\rangle =\hat{P}_{\text{G}}|\Phi _0\rangle =\prod_i\hat{P}%
_{i;\text{G}}|\Phi _0\rangle \;.  \label{1.5} 
\end{equation} 
Here, $|\Phi _0\rangle $ is any normalized single-particle product state and 
the local Gutzwiller projector $\hat{P}_{i;\text{G}}$ is defined as  
\begin{equation} 
\hat{P}_{i;\text{G}}=\prod_\Gamma \lambda _\Gamma ^{\hat{m}_\Gamma 
}=1+\sum_\Gamma \left( \lambda _\Gamma -1\right) \hat{m}_\Gamma \;. 
\label{1.6} 
\end{equation} 
To simplify our notation, we suppress the spatial indices wherever a 
misunderstanding is impossible, e.g., we omitted the index $i$ on the rhs. of 
eq. (\ref{1.6}). The real variational parameters $\lambda _\Gamma $ may have 
values between zero and one, where these two limits generate the 
ground-state both in the uncorrelated ($\hat{H}_{\text{at}}=0$) and the 
atomic limit ($\hat{H}_1=0$) of our Hamiltonian (\ref{1.1}). 
 
In Ref. \onlinecite{PRB98} we showed that the expectation value of the 
Hamiltonian (\ref{1.1}) can be evaluated for the wave-function (\ref{1.5}) 
in the limit of infinite spatial dimensions. In this section we only 
summarize the main ideas of the diagrammatic derivation which are important 
for our treatment of the spin-wave-problem in the next chapters. For all 
details we refer the reader to Ref. \onlinecite{PRB98}. 
 
First, let us consider the norm of the wave-function (\ref{1.5}),  
\begin{equation} 
\left\langle \Psi _{\text{G}}\,|\,\Psi _{\text{G}}\right\rangle 
=\prod_i\left\langle \Phi _0\left| \hat{P}_{i;\text{G}}^2\right| \Phi 
_0\right\rangle \;.  \label{1.7} 
\end{equation} 
The square of the local Gutzwiller-projector can be written as  
\begin{equation} 
\hat{P}_{i;\text{G}}^2=1+\sum_{I,I^{\prime }\,(|I|,|I^{\prime }|\geq 
2)}x_{i;I,I^{\prime }}\hat{n}_{i;I,I^{\prime }}^{\text{HF}}\;,  \label{1.8} 
\end{equation} 
where we introduced the (local) Hartree-Fock-operators  
\begin{mathletters} 
\begin{eqnarray} 
\hat{n}_{I,I}^{\text{HF}} &=&\prod_{\sigma \in I}\hat{n}_\sigma ^{\text{HF}%
}\;, \\[3pt] 
\hat{n}_\sigma ^{\text{HF}} &=&\hat{n}_\sigma ^{}-n_\sigma ^0 
\end{eqnarray} 
for $I=I^{\prime }$, and  
\begin{eqnarray} 
\hat{n}_{I,I^{\prime }}^{\text{HF}} &=&\biggl[\prod_{\sigma \in J}\hat{n}%
_\sigma ^{\text{HF}}\biggr]\hat{n}_{I_1,I_2}\qquad (J=I\cap I^{\prime 
};I=J\cup I_1;I^{\prime }=J\cup I_2)  \label{1.10} \\ 
\hat{n}_{I_1,I_2} &=&\prod_{\sigma _1\in I_1}\hat{c}_{\sigma 
_1}^{+}\prod_{\sigma _2\in I_2}\hat{c}_{\sigma _2}^{}  \label{1.10b} 
\end{eqnarray} 
for $I\neq I^{\prime }$. The operators $\hat{c}_{\sigma _1}^{+}$ ($\hat{c}%
_{\sigma _2}^{}$) in (\ref{1.10b}) should be placed in an ascending 
(descending) order. An explicit expression for the coefficients $%
x_{i;I,I^{\prime }}$ in (\ref{1.8}) is derived in Appendix (\ref{integr}). 
 
When we apply Wick's-Theorem to the right-hand-side of eq. (\ref{1.7}), all 
terms are represented by certain diagrams with lines  
\end{mathletters} 
\begin{equation} 
P_{i,j}^{\sigma ,\sigma ^{\prime }}=\left\langle \Phi _0\left| \hat{c}%
_{i,\sigma }^{+}\hat{c}_{j,\sigma ^{\prime }}^{}\right| \Phi _0\right\rangle  
\label{1.11} 
\end{equation} 
and local vertices $x_{i;I,I^{\prime }}$. The special form of the operator $%
\hat{P}_{i;\text{G}}^2$ in (\ref{1.8}) has two essential consequences for 
the structure of our diagrams. First, the definition of the 
Hartree-Fock operators together with eq. (\ref{1.1b}) guarantees that there 
are no local lines, i.e., 
\begin{equation} 
P_{i,i}^{\sigma ,\sigma ^{\prime }}=0\;. 
\end{equation} 
Second, the constraint $|I|,|I^{\prime }|\geq 2$ requires that at least four 
lines meet at every local vertex. When we evaluate the expectation values in  
\begin{mathletters} 
\label{1.12ges} 
\begin{eqnarray} 
\left\langle \hat{H}_{i;\text{at}}\right\rangle _{\Psi _{\text{G}}} 
&=&\sum_\Gamma E_\Gamma \left\langle \hat{m}_{i;\Gamma }\right\rangle _{\Psi 
_{\text{G}}}\text{ and }  \label{1.12} \\ 
\left\langle \hat{H}_1\right\rangle _{\Psi _{\text{G}}} &=&\sum_{i,j;\sigma 
,\sigma ^{\prime }}t_{i,j}^{\sigma ,\sigma ^{\prime }}\left\langle \hat{c}%
_{i;\sigma }^{+}\hat{c}_{j;\sigma ^{\prime }}^{\vphantom{+}}\right\rangle 
_{\Psi _{\text{G}}}  \label{1.12b} 
\end{eqnarray} 
we obtain diagrams, which contain one (for $\hat{H}_{i;\text{at}}$) or two 
(for $\hat{H}_1$) external vertices for the lattice sites $i$ (and $j$). If 
such a diagram possesses at least one internal vertex, we have lattice 
sites, which are connected by more than two lines. Such diagrams vanish in 
infinite dimensions and therefore we concluded in Ref. \onlinecite{PRB98} 
that the expectation values (\ref{1.12ges}) only include diagrams without 
any internal vertex. Thus, we can write the expectation values $\left\langle  
\hat{m}_\Gamma \right\rangle _{\Psi _{\text{G}}}=\left\langle \hat{m}%
_{i;\Gamma }\right\rangle _{\Psi _{\text{G}}}$ in (\ref{1.12}) as  
\end{mathletters} 
\begin{equation} 
m_\Gamma \equiv \left\langle \hat{m}_\Gamma \right\rangle _{\Psi _{\text{G}%
}}=\lambda _\Gamma ^2m_\Gamma ^0\equiv \lambda _\Gamma ^2\left\langle \hat{m}%
_\Gamma \right\rangle _0\; 
\end{equation} 
with uncorrelated expectation values $\left\langle ...\right\rangle _0\equiv 
\left\langle ...\right\rangle _{\Phi _0}$. This relation allows to replace 
the original variational parameters $\lambda _\Gamma $ by the new parameters  
$m_\Gamma $. The expectation value for a hopping term in (\ref{1.12b}) 
becomes  
\begin{mathletters} 
\label{1.14ges} 
\begin{eqnarray} 
\left\langle \hat{c}_{i;\sigma }^{+}\hat{c}_{j;\sigma ^{\prime }}^{%
\vphantom{+}}\right\rangle _{\Psi _{\text{G}}} &=&\sqrt{q_\sigma 
^{}q_{\sigma ^{\prime }}^{}}\left\langle \hat{c}_{i;\sigma }^{+}\hat{c}%
_{j;\sigma ^{\prime }}^{\vphantom{+}}\right\rangle _0\;  \label{1.14} \\ 
\sqrt{q_\sigma ^{}} &\equiv &\sqrt{\frac 1{n_\sigma ^0(1-n_\sigma ^0)}}%
\sum_{\Gamma ,\Gamma ^{\prime }}\sqrt{\frac{m_\Gamma m_{\Gamma ^{\prime }}}{%
m_\Gamma ^0m_{\Gamma ^{\prime }}^0}}  \label{qfac} \\ 
&&\times \sum_{{I,I}^{\prime }(\sigma \notin I,I^{\prime })}f_\sigma 
^I\,f_\sigma ^{I^{\prime }}\sqrt{m_{(I^{\prime }\cup \sigma )}^0m_{I^{\prime 
}}^0}T_{\Gamma ,(I\cup \sigma )}^{+}T_{(I^{\prime }\cup \sigma ),\Gamma 
}T_{\Gamma ^{\prime },I^{\prime }}^{+}T_{I,\Gamma ^{\prime }}.  \nonumber 
\end{eqnarray} 
Here, the fermionic sign function  
\end{mathletters} 
\begin{equation} 
f_\sigma ^I\equiv \langle I\cup \sigma |\hat{c}_\sigma ^{+}|I\rangle \; 
\end{equation} 
gives a minus (plus) sign if it takes an odd (even) number of 
anticommutations to shift the operator $\hat{c}_\sigma ^{+}$ to its proper 
place in the sequence of electron creation operators in $|I\cup \sigma 
\rangle $. Note that the numbers $q_\sigma ^{}$ in (\ref{1.14}) are just the 
diagonal-elements of the matrix $q_\sigma ^{\sigma ^{\prime }}$ introduced 
in Ref. \onlinecite{PRB98}, which is diagonal for our symmetry-restricted 
orbital basis (see Sec. \ref{sec2.1}).

\section{Spin Waves} 
\label{spin} 
The theoretical examination of spin-wave excitations requires the analysis 
of the imaginary part $\chi _T(\vec{q},E)$ of the transversal susceptibility%
\cite{Lovesey}, which is given as the retarded two-particle Greenfunction  
\begin{mathletters} 
\begin{eqnarray} 
G_T(\vec{q},E) &=&\frac 1L\langle \langle \hat{S}_{\vec{q}}^{+}\,;\hat{S}_{%
\vec{q}}^{-}\rangle \rangle _E  \label{3.1} \\ 
&=&-\frac iL\int_0^\infty dt\,e^{iEt}\left\langle \Psi _0\left| \left[ \hat{S%
}_{\vec{q}}^{+}(t),\hat{S}_{\vec{q}}^{-}(0)\right] \right| \Psi 
_0\right\rangle \;. 
\end{eqnarray} 
Here, we introduced the $\vec{q}$-dependent spin-flip operators  
\end{mathletters} 
\begin{mathletters} 
\begin{eqnarray} 
\hat{S}_{\vec{q}}^{+} &=&\sum_le^{i\vec{q}\vec{R}_l}\hat{S}%
_l^{+}=\sum_{l,b}e^{i\vec{q}\vec{R}_l}\hat{c}_{l,b,\uparrow }^{+}\hat{c}%
_{l,b,\downarrow }^{}  \label{3.2} \\ 
\hat{S}_{\vec{q}}^{-} &=&(\hat{S}_{\vec{q}}^{+})^{+}=\sum_{l,b}e^{-i\vec{q}%
\vec{R}_l}\hat{c}_{l,b,\downarrow }^{+}\hat{c}_{l,b,\uparrow }^{} 
\label{3.2b} 
\end{eqnarray} 
in the Heisenberg-picture, where the sum includes all ($L$) lattice sites $l$ 
and orbitals $b$. The magnetic excitations of the system are represented by 
poles of the Greenfunction $G_T(\vec{q},E)$ with energies $E>0$. For our 
further analysis we expand the ``spin-wave state''  
\end{mathletters} 
\begin{equation} 
\left| \Psi _{\vec{q}}^0\right\rangle \equiv \hat{S}_{\vec{q}}^{-}\left| 
\Psi _0\right\rangle  \label{3.3} 
\end{equation} 
in terms of exact energy-eigenstates  
\begin{eqnarray} 
\left| \Psi _{\vec{q}}^0\right\rangle &=&\sum_nW_n\left| \Psi 
_n\right\rangle \;,  \label{3.4} \\ 
\hat{H}\left| \Psi _n\right\rangle &=&E_n\left| \Psi _n\right\rangle \;. 
\end{eqnarray} 
The Lehmann-representation of (\ref{3.1}),  
\begin{equation} 
G_T(\vec{q},E)=-\frac iL\sum_n\left[ \frac{\mid \langle \Psi _n\mid \hat{S}_{%
\vec{q}}^{-}\mid \Psi _0\rangle \mid ^2}{E-(E_n-E_0)+i\delta }-\frac{\mid 
\langle \Psi _n\mid \hat{S}_{\vec{q}}^{+}\mid \Psi _0\rangle \mid ^2}{%
E+(E_n-E_0)+i\delta }\right]  \label{3.4c} 
\end{equation} 
shows that there are poles in $G_T(\vec{q},E)$ for the energies $E_n-E_0>0$  
with weights $\left| W_n\right| ^2$. 
 
In a ferromagnetic system the state $\mid \Psi _{\vec{q}=\vec{0}}^0\rangle $ 
is also a ground state of $\hat{H}$, since the operator $\hat{S}_{\vec{q}=%
\vec{0}}^{-}$ just flips a spin in the spin-multiplet of the ground-state $%
\left| \Psi _0\right\rangle $. Therefore, we can conclude that $G_T(\vec{0}%
,E)$ has one isolated pole for $E-E_0=0$. Now we consider finite, but small 
values of $\vec{q}$, and assume that the expansion (\ref{3.4}) is still 
dominated by a narrow distribution of low-energy states. This scenario 
explains the pronounced peak in $\chi _T(\vec{q},E)$ for small values of $E$ 
and $\left| \vec{q}\,\right| $, which is seen in experiments and interpreted 
as a spin-wave excitation (see, e.g., Ref. \onlinecite{Lowde}). Then, the 
spin-wave dispersion $E_{\vec{q}}$ can be identified as the position of this 
peak, and $E_{\vec{q}}$ is approximately determined by the first moment of 
the distribution $\left| W_n\right| ^2$,  
\begin{mathletters} 
\label{3.6ges} 
\begin{eqnarray} 
E_{\vec{q}} &=&\frac{\sum_nE_n\,\left| W_n\right| ^2}{\sum_n\left| 
W_n\right| ^2}-\frac{\left\langle \Psi _0\left| \hat{H}\right| \Psi 
_0\right\rangle }{\left\langle \Psi _0\mid \Psi _0\right\rangle } 
\label{3.5} \\ 
&=&\frac{\left\langle \Psi _0\left| \hat{S}_{\vec{q}}^{+}\hat{H}\hat{S}_{%
\vec{q}}^{-}\right| \Psi _0\right\rangle }{\left\langle \Psi _0\left| \hat{S}%
_{\vec{q}}^{+}\hat{S}_{\vec{q}}^{-}\right| \Psi _0\right\rangle }-\frac{%
\left\langle \Psi _0\left| \hat{H}\right| \Psi _0\right\rangle }{%
\left\langle \Psi _0\mid \Psi _0\right\rangle }\;.  \label{3.6} 
\end{eqnarray} 
It is still impossible to derive the spin-wave dispersion $E_{\vec{q}}$ from 
eq. (\ref{3.6ges}) since we do not know the ground-state $\left| \Psi 
_0\right\rangle $ of our multi-band Hamiltonian (\ref{1.1}). If we assume, 
however, that the variational wave function $\left| \Psi _{\text{G}%
}\right\rangle $ is a good approximation for the true ground-state $\left| 
\Psi _0\right\rangle $ we may substitute $\left| \Psi _0\right\rangle $ in 
eq. (\ref{3.6ges}) by the variational wave function $\left| \Psi _{\text{G}%
}\right\rangle $. 
 
In the next section we will evaluate the ``variational'' spin-wave 
dispersion  
\end{mathletters} 
\begin{equation} 
E_{\vec{q}}^{var}=\frac{\left\langle \Psi _{\text{G}}\left| \hat{S}_{\vec{q}%
}^{+}\hat{H}\hat{S}_{\vec{q}}^{-}\right| \Psi _{\text{G}}\right\rangle }{%
\left\langle \Psi _{\text{G}}\left| \hat{S}_{\vec{q}}^{+}\hat{S}_{\vec{q}%
}^{-}\right| \Psi _{\text{G}}\right\rangle }-\frac{\left\langle \Psi _{\text{%
G}}\left| \hat{H}\right| \Psi _{\text{G}}\right\rangle }{\left\langle \Psi _{%
\text{G}}\mid \Psi _{\text{G}}\right\rangle }  \label{3.7} 
\end{equation} 
in the limit of large spatial dimensions. It should be noted that this 
quantity obviously obeys no strict upper-bound properties. Nevertheless, we 
expect that $E_{\vec{q}}^{\exp }<E_{\vec{q}}^{var}$ is fulfilled since the 
expectation values (\ref{3.6ges}) includes high-energy states which do not 
belong to the spin-wave excitation seen in experiments. 
 
In principle, transversal spin-excitations are given as peaks both in $\chi 
_T(\vec{q},E)$ and $\chi _T(\vec{q},E_0-E)$ for energies $E>E_0$. In other 
word, we also had to consider the contributions from the Green function $%
\langle \langle \hat{S}_{\vec{q}}^{-}\,;\hat{S}_{\vec{q}}^{+}\rangle \rangle 
_E$ in our calculation. These contributions are identical to the second term 
in eq. (\ref{3.4c}) and we could include them by using the proper spin-wave 
state  
\begin{equation} 
\left| \tilde{\Psi}_{\vec{q}}^0\right\rangle \equiv \left( \hat{S}_{\vec{q}%
}^{-}+\hat{S}_{\vec{q}}^{+}\right) \left| \Psi _G\right\rangle   \label{3.8} 
\end{equation} 
in our variational approach. However, the contributions from the second 
operator in (\ref{3.8}) vanish for $\vec{q}=\vec{0}$ and may be neglected 
for small values of $|\vec{q}\,|$, where spin-wave excitations are observed 
in experiments. Nevertheless, there is no fundamental obstacle to extend our 
diagrammatic approach to the more general spin-wave state (\ref{3.8}).

\section{Variational Spin-Wave Dispersion} 
\label{calcu} 
\subsection{General considerations} 
 
In order to determine the variational spin-wave dispersion (\ref{3.7}) we 
need to examine the norm of the state $\left| \Psi _{\vec{q}}^{\text{G}%
}\right\rangle \equiv \hat{S}_{\vec{q}}^{-}\left| \Psi _{\text{G}%
}\right\rangle \;,$  
\begin{equation} 
N_{\vec{q}}\equiv \left\langle \Psi _{\text{G}}\left| \hat{S}_{\vec{q}}^{+}%
\hat{S}_{\vec{q}}^{-}\right| \Psi _{\text{G}}\right\rangle \;,  \label{4.1} 
\end{equation} 
and the expectation values  
\begin{mathletters} 
\label{4.2} 
\begin{eqnarray} 
\frac{\left\langle \Psi _{\vec{q}}^{\text{G}}\left| \hat{H}_{\text{at}%
}\right| \Psi _{\vec{q}}^{\text{G}}\right\rangle }{N_{\vec{q}}} 
&=&\sum_\Gamma E_\Gamma \sum_{i,j,k}e^{i\vec{q}(\vec{R}_i-\vec{R}_j)}\frac{%
\left\langle \Psi _{\text{G}}\left| \hat{S}_i^{+}\hat{m}_{k;\Gamma }\hat{S}%
_j^{-}\right| \Psi _{\text{G}}\right\rangle }{N_{\vec{q}}}\;,  \label{4.2a} 
\\ 
\frac{\left\langle \Psi _{\vec{q}}^{\text{G}}\left| \hat{H}_1\right| \Psi _{%
\vec{q}}^{\text{G}}\right\rangle }{N_{\vec{q}}} &=&\sum_{i,j,k,l}e^{i\vec{q}(%
\vec{R}_i-\vec{R}_j)}\sum_{\sigma _k,\sigma _l}t_{k,l}^{\sigma _k,\sigma _l}%
\frac{\left\langle \Psi _{\text{G}}\left| \hat{S}_i^{+}\hat{c}_{k;\sigma 
_k}^{+}\hat{c}_{l;\sigma _l}^{}\hat{S}_j^{-}\right| \Psi _{\text{G}%
}\right\rangle }{N_{\vec{q}}}\;.  \label{4.2b} 
\end{eqnarray} 
Before we start to evaluate these quantities, it is necessary to discuss two 
general problems. First, let us consider the norm  
\end{mathletters} 
\begin{equation} 
N_{\vec{q}}=\sum_{i,j,b,b^{\prime }}e^{i\vec{q}(\vec{R}_i-\vec{R}%
_j)}\left\langle \Psi _{\text{G}}\left| \hat{c}_{i,b,\uparrow }^{+}\hat{c}%
_{i,b,\downarrow }^{}\hat{c}_{j,b^{\prime },\downarrow }^{+}\hat{c}%
_{j,b^{\prime },\uparrow }^{}\right| \Psi _{\text{G}}\right\rangle \; 
\label{4.3} 
\end{equation} 
in the special case $\vec{q}=\vec{0}$, where $\hat{S}_{\vec{q}=\vec{0}}^{-}$ 
is just the total spin-flip operator $\hat{S}^{-}$. When we assume that $%
\left| \Psi _{\text{G}}\right\rangle $ has the correct spin-symmetry, i.e. 
it is an eigenstate of  
\begin{equation} 
\hat{S}^z=\sum_i\hat{S}_{i,z}\;\;\;\;\text{and}\;\;\;\;\Sq=\Bigl(\sum_i\vec{S%
}_i\Bigr)^2  \label{4.3b} 
\end{equation} 
with eigenvalues $S_{\text{G}}^z$ and $S_{\text{G}}^z(S_{\text{G}}^z+1)$, 
respectively, we obtain  
\begin{equation} 
N_{\vec{q}=\vec{0}}=2S_{\text{G}}^z\,\left\langle \Psi _{\text{G}}\mid \Psi 
_{\text{G}}\right\rangle \;.  \label{4.4} 
\end{equation} 
Here, we used the well-known equation  
\begin{equation} 
\hat{S}^{+}\hat{S}^{-}=\vec{S}^2-\hat{S}^z(\hat{S}^z-1)\;  \label{4.4b} 
\end{equation} 
for spin operators. In general, however, the wave functions $\left| \Psi _{%
\text{G}}\right\rangle $, as defined in eq. (\ref{1.6}), do not fulfill this 
symmetry. Therefore, it is necessary to introduce some additional 
constraints on our variational parameters $\lambda _\Gamma $ in (\ref{1.6}) 
to guarantee that $\left| \Psi _{\text{G}}\right\rangle $ is an eigenstate 
of $\Sq$. In Appendix (\ref{symm}) we explain how these relations may be 
chosen. 
 
The second problem arises from the evaluation of $\vec{q}$-dependent 
quantities in the limit of large spatial dimensions $D$. For example, let us 
consider the Hartree-Fock case, where $\left| \Psi _{\text{G}}\right\rangle 
=|\Phi _0\rangle $ is a spin-polarized one-particle state with $n_{\uparrow 
}>n_{\downarrow }$. We find  
\begin{mathletters} 
\label{4.5} 
\begin{eqnarray} 
\frac{N_{\vec{q}}^0}L &=&\frac 1L\sum_{b,b^{\prime }}\sum_{i,j}e^{i\vec{q}(%
\vec{R}_i-\vec{R}_j)}\left\langle \Phi _0\left| \hat{c}_{i,b,\uparrow }^{+}%
\hat{c}_{i,b,\downarrow }^{}\hat{c}_{j,b^{\prime },\downarrow }^{+}\hat{c}%
_{j,b^{\prime },\uparrow }^{}\right| \Phi _0\right\rangle   \label{4.5a} \\ 
&=&\sum_b\left( n_{b,\uparrow }^0-n_{b,\uparrow }^0n_{b,\downarrow }^0+\frac 
1L\sum_{i\neq j}e^{i\vec{q}(\vec{R}_i-\vec{R}_j)}P_{i,j}^{(b,\uparrow 
),(b,\uparrow )}P_{j,i}^{(b,\downarrow ),(b,\downarrow )}\right) \;, 
\label{4.5b} 
\end{eqnarray} 
where, for simplicity, we assumed that the expectation-values $%
P_{i,j}^{\sigma ,\sigma ^{\prime }}$ as defined in (\ref{1.11}) are diagonal 
with respect to the orbitals $b,b^{\prime }$. For any finite value of $\vec{q%
}$ the sum in eq. (\ref{4.5b}) vanishes as $1/D$. This means that the limits  
$\vec{q}\rightarrow 0$ and $D\rightarrow \infty $ do not commute, because  
\end{mathletters} 
\begin{equation} 
\lim_{\vec{q}\rightarrow \vec{0}}\lim_{D\rightarrow \infty }\frac{N_{\vec{q}%
}^0}L=\sum_bn_{b,\uparrow }^0\left( 1-n_{b,\downarrow }^0\right) \neq 
\sum_b\left( n_{b,\uparrow }^0-n_{b,\downarrow }^0\right) 
=\lim_{D\rightarrow \infty }\lim_{\vec{q}\rightarrow \vec{0}}\frac{N_{\vec{q}%
}^0}L\;.  \label{4.6} 
\end{equation} 
To overcome this problem we will evaluate expressions like the sum in (\ref 
{4.5b}) using the realistic three dimensional band-structure of our 
Hamiltonian (\ref{1.1}). This leads to results which are continuous in $\vec{%
q}$ and, consequently, reproduce the limit $\vec{q}=\vec{0}$ correctly . 
 
The norm of the state (\ref{4.1}) for a correlated wave function $\left| 
\Psi _{\text{G}}\right\rangle \neq |\Phi _0\rangle $ will contain diagrams 
of an arbitrary order $1/D^n$. In this paper we will only consider diagrams 
up to the leading order $n=1$. At first sight one may wonder whether or not $%
1/D$-terms need to be included in the ground-state energy expression as 
well. Fortunately, to order $1/D$, these diagrams only lead to a constant 
shift of the energy expectation-values for $\left| \Psi _{\text{G}%
}\right\rangle $ and $\left| \Psi _{\vec{q}}^{\text{G}}\right\rangle $. 
Thus, for our calculation we may neglect the $1/D$-contributions to the 
ground-state energy since we are only interested in the difference between 
these two energies. 
 
In the next subsection we evaluate the norm (\ref{4.3}). The derivation of 
the expectation values (\ref{4.2a}) and (\ref{4.2b}) requires no additional 
techniques. The cumbersome calculations are deferred to appendices \ref 
{interac} and \ref{kinetic}. 
 
\subsection{Evaluation of the norm} 
 
The norm of the state (\ref{4.1}) can be written as  
\begin{eqnarray} 
N_{\vec{q}} &=&\sum_i\prod_{l(\neq i)}\left\langle \left( \hat{P}_{i;\text{G}%
}\hat{S}_i^{+}\hat{S}_i^{-}\hat{P}_{i;\text{G}}\right) \,\hat{P}_{l;\text{G}%
}^2\right\rangle _{\Phi _0}  \label{4.7a} \\ 
&&+\sum_{i\neq j}e^{i\vec{q}(\vec{R}_i-\vec{R}_j)}\prod_{l(\neq 
i,j)}\left\langle \left( \hat{P}_{i;\text{G}}\hat{S}_i^{+}\hat{P}_{i;\text{G}%
}\right) \left( \hat{P}_{j;\text{G}}\hat{S}_j^{-}\hat{P}_{j;\text{G}}\right) 
\,\hat{P}_{l;\text{G}}^2\right\rangle _{\Phi _0}\;.  \nonumber 
\end{eqnarray} 
When we use eq. (\ref{1.8}) and apply Wick's theorem, we obtain diagrams 
with external vertices for the lattice sites $i$ and $j$ and internal 
vertices generated by the Hartree-Fock operators $x_{l;I,I^{\prime }}\hat{n}%
_{l;I,I^{\prime }}^{\text{HF}}$ in $\hat{P}_{l;\text{G}}^2$. In Refs. %
\onlinecite{BGW,PRB98} it was shown that we only have to evaluate the 
connected diagrams since the unconnected terms just give the norm of the 
Gutzwiller wave-function $N_{\text{G}}\equiv \left\langle \Psi _{\text{G}%
}\mid \Psi _{\text{G}}\right\rangle $. For the evaluation of the first line (%
\ref{4.7a}) we use the local relations  
\begin{mathletters} 
\label{4.8ges} 
\begin{eqnarray} 
\hat{S}^{+}\hat{S}^{-} &=&\sum_\Gamma S_{+}(\Gamma )\hat{m}_\Gamma \;, \\ 
S_{\pm }(\Gamma ) &\equiv &S(\Gamma )\left[ S(\Gamma )+1\right] -S_z(\Gamma 
)\left[ S_z(\Gamma )\mp 1\right] \; 
\end{eqnarray} 
which follow from eq. (\ref{4.4b}). Here, we introduced the total spin $%
S(\Gamma )$ and the spin-component $S_z(\Gamma )$ of the atomic eigenstates $%
\left| \Gamma \right\rangle $. Hence, the expectation value of $\hat{S}_i^{+}%
\hat{S}_i^{-}$ is given as a linear function of the variational parameters $%
m_\Gamma $ and we may write (\ref{4.7a}) as  
\end{mathletters} 
\begin{mathletters} 
\label{4.9ges} 
\begin{eqnarray} 
\frac{N_{\vec{q}}}{LN_{\text{G}}} &=&\sum_\Gamma S_{-}(\Gamma )m_\Gamma  
\label{4.9-a} \\ 
&&+\frac 1L\sum_{i\neq j}e^{i\vec{q}(\vec{R}_i-\vec{R}_j)}\prod_{l(\neq 
i,j)}\left\{ \left( \hat{P}_{i;\text{G}}\hat{S}_i^{+}\hat{P}_{i;\text{G}%
}\right) \left( \hat{P}_{j;\text{G}}\hat{S}_j^{-}\hat{P}_{j;\text{G}}\right)  
\hat{P}_{l;\text{G}}^2\right\} _{\Phi _0}^c,  \label{4.9} 
\end{eqnarray} 
where $\left\{ ...\right\} _{\Phi _0}^c$ denotes the application of Wick's 
theorem and taking into account only the connected diagrams. 
 
For a further analysis of (\ref{4.9}) we introduce indices $\frak{D}=(\sigma 
_1\sigma _2)$ for pairs of spin-orbit states $\sigma _i$ and the basic 
RPA-diagrams  
\end{mathletters} 
\begin{equation} 
\tilde{P}_{\frak{D}}^{\frak{D}^{\prime }}(\vec{q}\,)=\tilde{P}_{(\sigma 
_3\sigma _4)}^{(\sigma _1\sigma _2)}(\vec{q}\,)\equiv -\frac 1L\sum_{i\neq 
j}e^{i\vec{q}(\vec{R}_i-\vec{R}_j)}P_{i,j}^{\sigma _1\sigma 
_4}P_{j,i}^{\sigma _3\sigma _2}\;.  \label{4.10} 
\end{equation} 
$\tilde{P}_{\frak{D}}^{\frak{D}^{\prime }}(\vec{q}\,)$ can be evaluated in 
momentum space as  
\begin{equation} 
\tilde{P}_{\frak{D}}^{\frak{D}^{\prime }}(\vec{q}\,)=\delta _{\sigma 
_1}^{\sigma _4}\delta _{\sigma _3}^{\sigma _2}n_{\sigma _1}^0n_{\sigma 
_2}^0-\frac 1L\sum_{\vec{k}}n_{\vec{k}}^{\sigma _1\sigma _4}n_{\vec{k}+\vec{q%
}}^{\sigma _3\sigma _2}  \label{4.11} 
\end{equation} 
with expectation values  
\begin{equation} 
n_{\vec{k}}^{\sigma \sigma ^{\prime }}\equiv \left\langle \hat{c}_{\vec{k}%
\sigma }^{+}\hat{c}_{\vec{k}\sigma ^{\prime }}\right\rangle _{\Phi _0}\; 
\label{4.12} 
\end{equation} 
and a modified Kronecker-symbol $\delta _\sigma ^{\sigma ^{\prime }}=\delta 
_{\sigma ,\sigma ^{\prime }}^{}$. Note that, in contrast to the indices $I$%
,\ the order of the two spin-orbit states in $\frak{D}=(\sigma _1\sigma _2)$ 
is significant. Here, its first and second element specify a particle which 
enters or leaves a vertex, respectively. 
 
For large spatial dimensions (i.e., up to the order $1/D$), the only 
contributions in (\ref{4.9}) are RPA-type diagrams as shown in Fig (\ref 
{fig1}). The internal vertices $\tilde{x}_{\frak{D}}^{\frak{D}^{\prime }}$ 
are generated by the operators $\hat{P}_{l;\text{G}}^2$, which can be 
expressed in terms of Hartree-Fock operators $x_{l;I,I^{\prime }}\hat{n}%
_{l;I,I^{\prime }}^{\text{HF}}$ (see eq. (\ref{1.8})). For our RPA-diagrams 
we only have to consider two-fermion-operators in $\hat{n}_{l;I,I^{\prime 
}}^{\text{HF}}$. A general vertex with $n$ incoming and outgoing lines is 
determined by the operator  
\begin{eqnarray} 
&&\sum_{I_1,I_2(I_1\cap I_2=\emptyset )}\sum_{J(I_1,I_2\notin J)}x_{J\cup 
I_1,J\cup I_2}\left\{ ...\hat{n}_{J\cup I_1,J\cup I_2}^{\text{HF}%
}...\right\} _{\Phi _0}^c \\ 
&\rightarrow &\sum_{I_1,I_2(I_1\cap I_2=\emptyset )}\sum_{J(I_1,I_2\notin 
J)}y_{J\cup I_1,J\cup I_2}\left\{ ...\hat{n}_{J\cup I_1,J\cup 
I_2}...\right\} _{\Phi _0}^c\;,  \nonumber 
\end{eqnarray} 
with  
\[ 
y_{J\cup I_1,J\cup I_2}=x_{J\cup I_1,J\cup I_2}f_{I_1}^Jf_{I_2}^J\;.  
\] 
Thus, the internal vertex $\tilde{x}_{\frak{D}}^{\frak{D}^{\prime }}=\tilde{x%
}_{(\sigma _3\sigma _4)}^{(\sigma _1\sigma _2)}$ at lattice site $l$ stems 
from the term  
\begin{equation} 
y_{l;(\sigma _1,\sigma _3)(\sigma _2,\sigma _4)}\left\{ ...\left( \hat{c}%
_{l,\sigma _1}^{+}\hat{c}_{l,\sigma _2}^{}\right) \left( \hat{c}_{l,\sigma 
_3}^{+}\hat{c}_{l,\sigma _4}^{}\right) ...\right\} _{\Phi _0}^cf_{\sigma 
_1}^{\sigma _3}\,f_{\sigma _2}^{\sigma _4}\;.  \label{4.13} 
\end{equation} 
The round brackets indicate that Wick's theorem may not be applied to 
operators in different brackets. Thus, we can identify $\tilde{x}_{\frak{D}%
}^{\frak{D}^{\prime }}$ as  
\begin{equation} 
\tilde{x}_{\frak{D}}^{\frak{D}^{\prime }}=\tilde{x}_{(\sigma _3\sigma 
_4)}^{(\sigma _1\sigma _2)}=f_{\sigma _2}^{\sigma _4}\,f_{\sigma _1}^{\sigma 
_3}\,x_{(\sigma _1,\sigma _3),(\sigma _2,\sigma _4)}\;.  \label{4.14a} 
\end{equation} 
An explicit expression for $x_{I,I^{\prime }}$ is derived in Appendix \ref 
{integr}. 
 
Our infinite RPA-sum (see Fig \ref{fig1}) for the second term in (\ref{4.9}) 
leads to the following matrix-equation for $\tilde{\Omega}_{\frak{D}}^{\frak{%
D}^{\prime }}(\vec{q}\,)$,  
\begin{mathletters} 
\label{4.15ges} 
\begin{eqnarray} 
\tilde{\Omega}(\vec{q}\,) &\equiv &\tilde{P}(\vec{q}\,)+\tilde{P}(\vec{q}%
\,)\cdot \tilde{x}\cdot \tilde{P}(\vec{q}\,)+\tilde{P}(\vec{q}\,)\cdot  
\tilde{x}\cdot \tilde{P}(\vec{q}\,)\cdot \tilde{x}\cdot \tilde{P}(\vec{q}%
\,)+...  \label{4.15} \\ 
&=&\tilde{P}(\vec{q}\,)+\tilde{P}(\vec{q}\,)\cdot \tilde{x}\cdot \tilde{%
\Omega}(\vec{q}\,)\;, 
\end{eqnarray} 
which has the solution  
\end{mathletters} 
\begin{equation} 
\tilde{\Omega}(\vec{q}\,)=(\hat{1}-\tilde{P}(\vec{q}\,)\cdot \tilde{x}%
)^{-1}\cdot \tilde{P}(\vec{q}\,)\;.  \label{4.16} 
\end{equation} 
Here, the dot ``$\,\cdot \,$''indicates the usual product between two 
matrices. 
 
Finally, we have to examine the external vertices, generated by the 
spin-operators $\hat{S}_i^{+}$ and $\hat{S}_j^{-}$ in (\ref{4.9}). First, we 
find for local operators  
\begin{mathletters} 
\label{4.17ges} 
\begin{eqnarray} 
\hat{P}_{\text{G}}S^{+}\hat{P}_{\text{G}} &=&\sum_{\Gamma ^{\prime },\Gamma 
}\lambda _{\Gamma ^{\prime }}\lambda _\Gamma \left| \Gamma ^{\prime 
}\right\rangle \left\langle \Gamma ^{\prime }\right| S^{+}\left| \Gamma 
\right\rangle \left\langle \Gamma \right|  \\ 
&=&\sum_\Gamma \lambda _{\Gamma _{+}}\lambda _\Gamma \sqrt{S_{+}(\Gamma )}%
\left| \Gamma _{+}\right\rangle \left\langle \Gamma \right| \;, 
\label{4.17b} 
\end{eqnarray} 
where the normalized spin-flip-state  
\end{mathletters} 
\begin{equation} 
\left| \Gamma _{+}\right\rangle \equiv \frac 1{\sqrt{S_{+}(\Gamma )}}\hat{%
S}^{+}\left| \Gamma \right\rangle \;  \label{4.18a} 
\end{equation} 
was introduced. We may write eq. (\ref{4.17b}) as  
\begin{equation} 
\hat{P}_{\text{G}}S^{+}\hat{P}_{\text{G}}=\sum_\Gamma \lambda _{\Gamma 
_{+}}\lambda _\Gamma \sqrt{S_{+}(\Gamma )}\sum_{I_1,I_2}T_{I_1,\Gamma 
_{+}}^{}T_{\Gamma ,I_2}^{+}\hat{m}_{I_1,I_2}  \label{4.18b} 
\end{equation} 
with  
\begin{mathletters} 
\label{4.18ges} 
\begin{eqnarray} 
\hat{m}_{I_1,I_2} &=&f_I^Jf_{I^{\prime }}^J\hat{m}_J^{I,I^{\prime }}\,\hat{n}%
_{I,I^{\prime }}\;,  \label{4.18c} \\ 
\hat{m}_J^{I,I^{\prime }} &\equiv &\prod_{\sigma \in J}\hat{n}_\sigma 
\prod_{\sigma \in \bar{J}\backslash (I\cup I^{\prime })}\left( 1-\hat{n}%
_\sigma \right)   \label{4.18d} 
\end{eqnarray} 
and $J=I_1\cap I_2$, $I_1=J\cup I$, $I_2=J\cup I^{\prime }$. 
 
The external vertices in our RPA-diagrams are met by just two lines. 
Therefore, we only need the following one-particle contribution of the 
operator $\hat{m}_{I_1,I_2}$ in (\ref{4.18b}),  
\end{mathletters} 
\begin{eqnarray} 
\left\{ ...\hat{m}_{I_1,I_2}...\right\} _{\Phi _0}^C &\rightarrow 
&\sum_{\sigma _1,\sigma _2}{\frak{V}}_{_{I_1,I_2}}^{(\sigma _1\sigma 
_2)}\left\{ ...\hat{c}_{\sigma _1}^{+}\hat{c}_{\sigma _2}^{}...\right\} 
_{\Phi _0}^c  \label{4.20} \\ 
{\frak{V}}_{_{I_1,I_2}}^{(\sigma _1\sigma _2)} &\equiv &\sum_{J(\sigma 
_1,\sigma _2\notin J)}f_{\sigma _1}^J\,f_{\sigma _2}^Jm_{^J}^0\Bigl%
(\prod_{\sigma \in (\sigma _1,\sigma _2)}\frac 1{(1-n_\sigma ^0)}\Bigr%
)\left[ \delta _{I_1}^{J\cup \sigma _1}\delta _{I_2,}^{J\cup \sigma 
_2}-\delta _{\sigma _1}^{\sigma _2}\delta _{I_1}^J\delta _{I_2}^J\right] \;  
\nonumber 
\end{eqnarray} 
where we assumed that $\left| I_1\right| =\left| I_2\right| $. When we apply 
this result to the contribution of the external vertex (\ref{4.9}), we only 
need to consider the first term with $\sigma _1\neq \sigma _2$, because $%
\sigma _1,\sigma _2$ must have different spins. Thus, the expression for the 
external vertex, which stems from $\hat{S}_i^{+}$ becomes  
\begin{mathletters} 
\label{4.21ges} 
\begin{eqnarray} 
S_{(\sigma _1\sigma _2)}^{+} &=&\sum_\Gamma \lambda _{\Gamma _{+}}\lambda 
_\Gamma \sqrt{S_{+}(\Gamma )}\sum_{I_1,I_2}T_{I_1,\Gamma 
_{+}}^{}T_{\Gamma ,I_2}^{+}{\frak{V}}_{_{I_1,I_2}}^{(\sigma _1\sigma _2)} 
\label{4.21} \\ 
&=&\sum_\Gamma \lambda _{\Gamma _{+}}\lambda _\Gamma \sqrt{S_{+}(\Gamma )}%
\sum_{J(\sigma _1,\sigma _2\notin J)}T_{J\cup \sigma _1,\Gamma 
_{+}}^{}T_{\Gamma ,J\cup \sigma _2}^{+}\frac{f_{\sigma _1}^J\,f_{\sigma 
_2}^Jm_{^J}^0}{(1-n_{\sigma _1}^0)(1-n_{\sigma _2}^0)}\;.  \label{4.21b} 
\end{eqnarray} 
The expansion  
\end{mathletters} 
\begin{mathletters} 
\label{4.22ges} 
\begin{eqnarray} 
\left\{ ...\hat{P}_{\text{G}}\hat{S}^{-}\hat{P}_{\text{G}}...\right\} _{\Phi 
_0}^C &\rightarrow &\sum_{\sigma _1,\sigma _2}\left[ S_{(\sigma _1\sigma 
_2)}^{+}\left\{ ...\left( \hat{c}_{\sigma _1}^{+}\hat{c}_{\sigma 
_2}^{}\right) ...\right\} _{\Phi _0}^c\right] ^{*} \\ 
&=&\sum_{\sigma _1,\sigma _2}\left( S_{(\sigma _1\sigma _2)}^{+}\right) 
^{*}\left\{ ...\left( \hat{c}_{\sigma _2}^{+}\hat{c}_{\sigma _1}^{}\right) 
...\right\} _{\Phi _0}^c\;, 
\end{eqnarray} 
shows that the vertex-factor for the operator $\hat{S}_j^{-}$ in (\ref{4.9}) 
is given as  
\end{mathletters} 
\begin{equation} 
S_{(\sigma _1\sigma _2)}^{-}=\left( S_{(\sigma _2\sigma _1)}^{+}\right) 
^{*}\;.  \label{4.22} 
\end{equation} 
When we consider $S_{\frak{D}}^{+}$ and $S_{\frak{D}}^{-}$ as components of 
vectors $\vec{S}^{+}$ and $\vec{S}^{-}$ with respect to the indices $\frak{D} 
$ we obtain the following final result for the norm $N_{\vec{q}}$  
\begin{mathletters} 
\label{4.23ges} 
\begin{eqnarray} 
\frac{N_{\vec{q}}}{LN_{\text{G}}} &=&\sum_\Gamma S_{-}(\Gamma )m_\Gamma 
+\sum_{\frak{D}_1,\frak{D}_2}S_{\frak{D}_1}^{+}\,\tilde{\Omega}_{\frak{D}%
_2}^{\frak{D}_1}(\vec{q}\,)\,S_{\frak{D}_2}^{-}  \label{4.23d} \\ 
&=&\sum_\Gamma S_{-}(\Gamma )m_\Gamma +\vec{S}^{+}\cdot \,\tilde{\Omega}(%
\vec{q}\,)\cdot \,\vec{S}^{-}\;. 
\end{eqnarray} 
Note that the norm $N_{\text{G}}$ in (\ref{4.23d}) will cancel out when we 
calculate the expectation values (\ref{4.2}). 
 
A similar derivation gives us the expectation values (\ref{4.2}) of the 
kinetic energy and the Coulomb-interaction, see appendices \ref{interac} and  
\ref{kinetic}. There, the number of contributing diagrams is much larger 
(e.g., $18$ for the kinetic energy) since we have up to four external 
vertices. Nevertheless, the numerical evaluation of these terms is a minor 
technical problem as soon as the wave-function $\left| \Psi _{\text{G}%
}\right\rangle $ has been determined by the minimization of our variational 
energy expression. First studies show that the application of our 
variational scheme to iron and nickel represents a solvable numerical task  
\cite{WGB}. Further work in this direction is in progress. 
\end{mathletters}

\section{Application to a two-band model} 
\label{zweiband} 
In this chapter we will present the results for the spin-wave properties in 
a system with two degenerate $e_g$-type orbitals per lattice site. The 
appearance of ferromagnetism in this model has already been discussed in 
Ref. \onlinecite{PRB98}, both for the Hartree-Fock and the Gutzwiller 
theory. Thus, we will only summarize these results here, before we start to 
consider the spin-wave dispersion. For all details, we refer to Ref. %
\onlinecite{PRB98}. 
 
\subsection{Ferromagnetic properties} 
 
In a system with two degenerate orbitals the general atomic Hamiltonian (\ref 
{1.2}) becomes  
\begin{eqnarray} 
\hat{H}_{\text{at}} &=&U\sum_b\hat{n}_{b,\uparrow }\hat{n}_{b,\downarrow 
}+U^{\prime }\sum_{\sigma ,\sigma ^{\prime }}\hat{n}_{1,\sigma }\hat{n}%
_{2,\sigma ^{\prime }}-J\sum_\sigma \hat{n}_{1,\sigma }\hat{n}_{2,\sigma } \\%
[3pt] 
&&+J\sum_\sigma \hat{c}_{1,\sigma }^{+}\hat{c}_{2,-\sigma }^{+}\hat{c}%
_{1,-\sigma }^{\vphantom{+}}\hat{c}_{2,\sigma }^{\vphantom{+}}+J_{\text{C}%
}\Bigl(\hat{c}_{1,\uparrow }^{+}\hat{c}_{1,\downarrow }^{+}\hat{c}%
_{2,\downarrow }^{\vphantom{+}}\hat{c}_{2,\uparrow }^{\vphantom{+}}+\hat{c}%
_{2,\uparrow }^{+}\hat{c}_{2,\downarrow }^{+}\hat{c}_{1,\downarrow }^{%
\vphantom{+}}\hat{c}_{1,\uparrow }^{\vphantom{+}}\Bigr)\;.  \nonumber 
\end{eqnarray} 
In cubic symmetry the Coulomb- and exchange-integrals $U$, $U^{\prime }$, $J$ 
and $J_{\text{C}}$ are not independent from each other. Instead we have only 
two free parameters, because the relations $J=J_{\text{C}}$ and $U-U^{\prime 
}=2J$ hold. For the determination of the optimal variational wave function (%
\ref{1.5}), we need the $16$ eigenstates of the atomic Hamiltonian. The 
latter can be found in table I of Ref. \onlinecite{PRB98}. 
 
In the one-particle Hamiltonian $\hat{H}_1$ we take into account first and 
second nearest neighbor hopping matrix elements. Furthermore, we apply the 
two-center approximation for the hopping matrix elements, which are chosen 
as $t_{dd\sigma }^{(1)}=1\,$eV, $t_{dd\sigma }^{(2)}=0.25\,\text{eV}$, and $%
t_{dd\sigma }^{(1),(2)}:t_{dd\pi }^{(1),(2)}:t_{dd\delta 
}^{(1),(2)}=1:(-0.3):0.1$ (see Ref. \onlinecite{SlaterKoster} for the 
notation). The density of states for these parameters (see fig. 1 of Ref. %
\onlinecite{PRB98}) has a pronounced peak for the particle-density $n_\sigma 
\approx 0.3$. For this band-filling we observe the strongest tendency to 
generate a ferromagnetic order, in qualitative agreement with the simple 
Stoner criterion. Therefore, we consider the spin-wave properties of our 
system only for this optimum band-filling. The numerical evaluation in Ref. %
\onlinecite{PRB98} did not respect the global spin-symmetry of the 
wave-function $\left| \Psi _{\text{G}}\right\rangle $. In this work, we 
include the additional constraints on the variational-parameters $\lambda 
_\Gamma $ as described in Appendix \ref{symm}, and we obtain almost the same 
state as in Ref. \onlinecite{PRB98}. In other words, in the two-band-model 
even the general variational ground-state $\left| \Psi _{\text{G}%
}\right\rangle $, as defined in (\ref{1.5}), is a nearly perfect eigenstate 
of the global spin-operator. However, it is not clear so far, if this 
statement also holds for a more general multi-band model. 
 
In Ref. \onlinecite{PRB98} we found significant differences between the 
Hartree-Fock and the Gutzwiller theory. These differences have to be 
interpreted as a failure of the Hartree-Fock-theory since the variational 
space of the Hartree-Fock theory is included in our general class of 
Gutzwiller wave-functions. Although this statements holds for the 
corresponding RPA-theory as well, this theory is generally considered as the standard method 
in the context of spin-waves in itinerant ferromagnets. \cite{Cook} 
 
The ferromagnetic phase diagrams for our model is shown in Fig. \ref{fig3}. 
In the Hartree-Fock theory ferromagnetism occurs for considerably smaller 
values of the correlation parameters. The most important difference lies in 
the role of the interatomic exchange $J$. In the Gutzwiller theory a 
ferromagnetic ground-state exists only for finite values of $J$. This is 
completely different in the Hartree-Fock theory. There, the only relevant 
quantities are the Stoner-parameter $I=\frac{U+J}2$ and the density of 
states at the Fermi-level ${\frak{D}}_0(E_F)$ which enter the Stoner 
criterion  
\begin{equation} 
I{\frak{D}}_0(E_F)>1\;.  \label{zw1} 
\end{equation} 
This means that even for $J=0$ finite values of $U$ exist, where the system 
makes a transition into a ferromagnetic state. 
 
Another important difference between both theories occurs when we compare 
the condensation energies $E_{\text{cond}}$, i.e., the differences between 
the energies in the paramagnetic and the ferromagnetic ground states. This 
quantity should have the same order of magnitude as the Curie-temperature $%
T_{\text{C}}$ in itinerant ferromagnets. In Stoner theory we observe that $%
E_{\text{cond}}$ is typically of the order of $U$ and therefore much larger 
than $T_{\text{C}}$. This is in agreement with the general observation that 
mean field methods overestimate the stability of magnetic order. On the 
other hand, in our correlated electron approach we find relatively small 
values for the condensation energy even for interaction parameters as large 
as twice the bandwidth. 
 
\subsection{Spin-waves} 
 
Fig. \ref{fig2} shows the spin-wave dispersion in $(100)$-direction for four 
different magnetizations. As the lattice constant of our simple-cubic 
lattice we chose $a=2.5\AA $, the next neighbor distance in Nickel. Our 
Gutzwiller theory shows that the dispersion strongly depends on the 
magnetization, especially for small values of $\vec{q}$. The line shows the 
respective fits for the function  
\begin{equation} 
E_{\vec{q}}=Dq^2(1-\beta q^2)\;.  \label{zw2} 
\end{equation} 
Note that experimental results of the region where quartic corrections 
become dominant are usually not very accurate, since Stoner excitations 
reduce the lifetime of spin-waves significantly. The values for the 
spin-wave stiffness $D=1.4\rm{eV\AA ^2}$ and $D=1.2\rm{eV\AA ^2}$ in both cases of 
almost fully polarized magnetizations, $m=0.26$ and $m=0.28$, respectively, 
are of the right order of magnitude for Nickel where $D=0.43eV\AA ^2$. 
 
The spin-wave dispersion is almost isotropic, as can be seen for example in the inset of Fig. \ref{fig2}, where the dispersion is shown in the directions $(100)$ and $%
(110)$. Such an isotropic behaviour was also observed in experiments on iron 
and nickel and it is actually somewhat surprising, since the band-structure 
in these materials is far from being isotropic. However, we should have in 
mind that even in a metallic system local moments are formed due to the electrons' correlations. Therefore, spin-excitations may be interpreted as spin-fluctuations in a system with 
localized spins. In such a system we have a generic isotropy when the 
exchange coupling is dominated by terms between nearest neighbors.  
 
Our results show that the magnetic excitations in strong itinerant 
ferromagnets behave very similar to those in systems of localized spins. 
These low-energy excitations are responsible for the magnetic phase 
transition which occurs for temperatures much smaller than the typical 
Fermi energies in itinerant electron systems. This observation is consistent with 
the small condensation energy in our variational approach. To support this 
statement let us consider a Heisenberg-model  
\[ 
\hat{H}_S=-J\sum_{\langle i,j\rangle }\vec{S}_i\vec{S}_j 
\] 
on a cubic lattice. In this system we have a spin-wave stiffness $D=2SJa^2$. 
The value of the effective local moment in our itinerant system is given as $%
S\approx 0.6$ (see Ref. \onlinecite{PRB98}). Therefore we obtain $J=\frac 
D{2Sa^2}\approx 0.16\rm{eV}$. For an estimate of the Curie-temperature we use the 
results from quantum Monte-Carlo calculations $T_C=1.44JS^2$.\cite{Chen} In this way we 
find $T_C\approx 8\cdot 10^2K$ which is the same order of magnitude as the 
condensation energy $E_{\text{cond}}\approx 5\cdot 10^2K$. Thus, we can 
summarize that our variational approach gives a consistent picture of both 
of magnetic excitations and ground-state properties in strong itinerant ferromagnets. 
 
When the spin-waves are treated as non-interacting bosons, their 
contribution to the specific heat and the magnetization $M(T)$ may be 
calculated from $E_{\vec{q}}$. The first order contribution stems from the 
quadratic term and is the well-known $T^{3/2}$-law.\cite{Bloch} The next order depends 
on the behaviour of $E_{\vec{q}}$ for larger values of $q$. It is still not 
clear, wether or not the quartic term in (\ref{zw2}) describes the generic feature in 
the experiments. Some experiments indicate that the second term in $M(T)$ is  
$\alpha T^2$, which could be explained with a linear behaviour in $E_{\vec{q}%
}$. Our dispersion $E_{\vec{q}}$ also allows to calculate such 
temperature-dependent quantities numerically. However, we did not analyze 
this for our two-band model, because these results could not be compared to 
experiments.

\section{Summary and Conclusion} 
\label{conc} 
 
In this paper we presented a variational method for the description of 
spin-wave excitations in itinerant ferromagnets. Our starting point was a 
general multi-band Hubbard-model which is an appropriate model for elements 
of the iron group. Earlier work showed that ferromagnetism in metals 
requires substantial interaction strengths even in systems with orbital 
degeneracy. Therefore, we suspect that weak-coupling theories are not able 
to describe the physics of itinerant ferromagnets correctly. Our method is 
based on a variational study of multi-band Hubbard models with the help of 
generalized Gutzwiller wave functions. These wave functions yield the exact 
ground-state both in the uncorrelated and the atomic limit of the Hubbard 
model. Therefore, we expect them to describe reasonably the ground-state 
properties for finite values of the correlation parameters. 
 
From a theoretical point of view, spin waves are given as peaks in the 
imaginary part $\chi _{T}(\vec{q},E)$ of the transversal spin susceptibility 
. This quantity can be measured using inelastic neutron scattering. For $%
\vec{q}=\vec{0}$ the spin-symmetry of the ferromagnetic ground-state leads 
to an isolated peak $\delta (E)$ in $\chi _{T}$. We assume that this peak is 
broadened only moderately for finite values of $\vec{q}$. Then, the position 
of the peak, which is interpreted as the spin-wave dispersion $E_{\vec{q}}$, 
can be calculated as a static expectation-value of spin-operators in the 
ground state of the system. In our approximation we use the variational 
instead of the true ground state and calculate all expectation values in the 
limit of large spatial dimensions. 
 
Our results may be evaluated numerically for general multi-band models to 
describe iron and nickel. However, the main numerical problem is the 
determination of the optimum variational wave-function for these realistic 
systems with a non-trivial atomic Hamiltonian. Work in this direction is 
still in progress. In this work we applied our method to a two-band model. 
Here, we found a behaviour which is in qualitative agreement with 
experimental results for strong ferromagnets. The spin-wave dispersion $E_{%
\vec{q}}$ is almost isotropic and quadratic for small values of $\vec{q}$. 
For larger $\vec{q}$ we found quartic corrections, which are also seen in 
some of the experiments. The values of the spin-wave stiffness have the 
right order of magnitude compared to experiments. We concluded that the 
low-lying magnetic excitations in our correlated and itinerant electron 
system are similar to those in a localized spin systems. When we estimate 
the Curie-temperature $T_C$ from our spin-wave properties and compare it to 
the condensation energy we find a consistent picture in our variational approach. The ferromagnetic 
phase transition is driven by spin-waves and the value for $T_C$ is 
therefore much smaller than typical Fermi energies in itinerant electron systems. 
 
\section{Acknowledgement}

The author is very grateful to F. Gebhard for a critical reading of the manuscript and many helpful discussions.

\appendix

\section{Global spin-symmetry} 
\label{symm} 
The variational spin-wave dispersion (\ref{3.7}) is gapless only if the 
variational wave-function $|\Psi _{\text{G}}\rangle $ is an eigenstate of 
the global spin-operator (see eq. (\ref{4.3b})). This symmetry is not 
fulfilled for an arbitrary choice of the variational parameters $\lambda 
_\Gamma $ in (\ref{1.6}). In this appendix, we present two ways to implement 
this symmetry. 
 
The first way to guarantee the global spin symmetry of $|\Psi _{\text{G}%
}\rangle $ starts from the fact that the one-particle product-state $|\Phi 
_0\rangle $ in (\ref{1.5}) may in general be chosen as an eigenstate both of  
$\Sq$ and $\hat{S}^z$. In this case, $|\Psi _{\text{G}}\rangle $ is an 
eigenstate of $\hat{S}^z$ if the states $|\Gamma \rangle $ are chosen as 
eigenstates of the local spin-operator in $z$-direction. Then, the correct 
spin-symmetry of $|\Psi _{\text{G}}\rangle $ is ensured if $|\Psi _{\text{G}%
}\rangle $ is an eigenstate of ${\hat{S}}^{+}{\hat{S}}^{-}$. It is 
equivalent to demand that  
\begin{equation} 
\Bigl\langle {\hat{S}}^{+}{\hat{S}}^{-}\Bigr\rangle _{\Psi _{\text{G}}}=%
\frac{2S_z}{N_{\text{G}}}\;,  \label{f1} 
\end{equation} 
since we may assume that the spin in $|\Psi _{\text{G}}\rangle $ has a 
maximal component in $z$-direction. The left-hand-side of eq. (\ref{f1}) is 
just the norm $N_{\vec{q}}$ for $\vec{q}=\vec{0}$ which was derived in (\ref 
{4.23d}). Thus, (\ref{f1}) together with (\ref{4.23d}) leads to {\sl one} 
additional condition for the variational-parameters $\lambda _\Gamma $. 
However, this condition is not very helpful, because equation (\ref{4.23d}) 
includes the optimum values $\lambda _\Gamma $. This means that (\ref{f1}) 
has to be included in the minimization algorithm, which determines the 
optimum wave-function $|\Psi _{\text{G}}\rangle $; numerically, this is a 
very difficult problem. 
 
In the following we propose a second, more feasible strategy for the 
implementation of the correct spin-symmetry. To this end, we arrange the 
orbitals on the atoms of our system into groups which carry the index of the 
respective representation $D$ of the point-symmetry group. Then, we define 
the operators  
\begin{equation} 
\hat{N}_D^s=\sum_{i,b\in D}\hat{n}_{i;(bs)}\;,\;\;\;\;\hat{M}%
_D^s=\sum_{i,b\in D}\hat{m}_{i;(bs)}\;, 
\end{equation} 
for the gross and net number of electrons in orbitals of the representation $%
D$ and with spin~$s$. Again, we assume that each representation occurs only 
once. Under this condition, group theoretical arguments show that the 
following relation holds  
\begin{equation} 
\sum_{b\in D}\hat{m}_{i;(bs)}=\sum_{b\in D}\hat{n}_{i;(bs)}-\sum_{\Gamma 
(|\Gamma |\geq 2)}f_D^s(\Gamma )\hat{m}_{i;\Gamma }\;,  \label{f3} 
\end{equation} 
where  
\begin{equation} 
f_D^s(\Gamma )=\sum_{b\in D}\sum_{I[(b,s)\notin I]}\left| T_{\Gamma ,I\cup 
(b,s)}\right| ^2\;. 
\end{equation} 
Thus, we can write  
\begin{mathletters} 
\begin{eqnarray} 
\hat{M}_D^s &=&\hat{N}_D^s-\sum_{\Gamma (|\Gamma |\geq 2)}f_D^s(\Gamma )\hat{%
M}_\Gamma \;, \\ 
\hat{M}_\Gamma &\equiv &\sum_i\hat{m}_{i;\Gamma }\;, 
\end{eqnarray} 
and the Gutzwiller-projector (\ref{1.6}) becomes  
\end{mathletters} 
\begin{equation} 
\hat{P}_{\text{G}}=\prod_{D,s}\left( \lambda _D^s\right) ^{\hat{N}%
_D^s}\prod_\Gamma \tilde{\lambda}_\Gamma ^{\hat{M}_\Gamma }\;. 
\end{equation} 
Here, we already assumed that the parameters $\lambda _{(b,s)}$ ($\equiv 
\lambda _D^s$) are the same for all orbitals, which belong to $D$. Further, 
we introduced  
\begin{equation} 
\tilde{\lambda}_\Gamma =\lambda _\Gamma \cdot \prod_{D,s}\left( \lambda 
_D^s\right) ^{-f_D^s(\Gamma )}\;.  \label{f7} 
\end{equation} 
Now, we postulate the following conditions which ensure that $|\Psi _{\text{G%
}}\rangle $ is an eigenstate of the operator $\Sq$: 
 
\begin{itemize} 
\item[(I):]  $\lambda _D^s=\lambda _{D^{\prime }}^s\equiv \lambda _s$ for 
all representations $D,D^{\prime }$. 
 
\item[(II):]  For all states $\left| \Gamma _S^{S_z}\right\rangle ,\left| 
\Gamma _S^{S_z^{\prime }}\right\rangle $, which belong to the same 
spin-multiplet ($\equiv \Gamma ^S$) with spin $S$, we have $\tilde{\lambda}%
_{\Gamma _S^{S_z}}=\tilde{\lambda}_{\Gamma _S^{S_z^{\prime }}}\equiv \tilde{%
\lambda}_{\Gamma _S}$. 
\end{itemize} 
 
For the prove of this statement, we can first conclude from (I) that the 
state $|\Phi _0\rangle $ is an eigenstate of  
\begin{equation} 
\prod_{D,s}\left( \lambda _D^s\right) ^{\hat{N}_D^s}=\prod_s\lambda _s^{\hat{%
N}_s}\;, 
\end{equation} 
where $\hat{N}_s$ is the number-operator for electrons with spin $s$. Thus, 
we have  
\begin{mathletters} 
\begin{eqnarray} 
{\hat{S}}^{+}{\hat{S}}^{-}|\Psi _{\text{G}}\rangle &=&{\hat{S}}^{+}{\hat{S}}%
^{-}\prod_s\lambda _s^{\hat{N}_s}\prod_\Gamma \tilde{\lambda}_\Gamma ^{\hat{M%
}_\Gamma }|\Phi _0\rangle \\ 
&\sim &{\hat{S}}^{+}{\hat{S}}^{-}\prod_\Gamma \tilde{\lambda}_\Gamma ^{\hat{M%
}_\Gamma }|\Phi _0\rangle \;, 
\end{eqnarray} 
since $\left[ \hat{N}_s,\hat{M}_\Gamma \right] =0$ for all $s,\Gamma $. We 
introduce the operator  
\end{mathletters} 
\begin{equation} 
\hat{M}_{\Gamma _S}\equiv \sum_{S_z=-S}^S\hat{M}_{\Gamma _S^{S_z}}\;, 
\end{equation} 
which has the property $\left[ {\hat{S}}^{\pm },\hat{M}_{\Gamma _S}\right] 
=0 $. Then, condition (II) finally gives  
\begin{eqnarray} 
{\hat{S}}^{+}{\hat{S}}^{-}|\Psi _{\text{G}}\rangle &\sim &{\hat{S}}^{+}{\hat{%
S}}^{-}\prod_{\Gamma _S}\tilde{\lambda}_{\Gamma _S}^{\hat{M}_{\Gamma 
_S}}|\Phi _0\rangle \\ 
&\sim &\prod_{\Gamma ^S}\tilde{\lambda}_{\Gamma _S}^{\hat{M}_{\Gamma _S}}{%
\hat{S}}^{+}{\hat{S}}^{-}|\Phi _0\rangle \sim |\Psi _{\text{G}}\rangle \;  
\nonumber 
\end{eqnarray} 
such that $|\Psi _{\text{G}}\rangle $ is an eigenstate of $\Sq$. 
 
Note, that condition (II) reduces the number of variational-parameters 
significantly, since all parameters $\lambda _\Gamma $ for states $\left| 
\Gamma \right\rangle $, which belong to the same spin-multiplet $\Gamma _S$ 
are now determined by just one parameter $\tilde{\lambda}_{\Gamma _S}$. 
However, this restriction still allows different occupations of the several $%
S_z$ components, since  
\begin{equation} 
m_{\Gamma _S^{S_z}}=\lambda _{\Gamma _S}m_{\Gamma _S^{S_z}}^0  \label{f12} 
\end{equation} 
also depends on the one-particle-state $|\Phi _0\rangle $. 
 
The one-particle-occupations  
\begin{equation} 
m_{(bs)}=\lambda _{(bs)}m_{(bs)}^0=\lambda _sm_{(bs)}^0\equiv m_D^s\;, 
\label{f13} 
\end{equation} 
\ with $b\in D$ depends on the quantities $n_D^s\equiv n_{(bs)}$ and $%
m_{\Gamma _S^{S_z}}$ (see eq. (\ref{f3})),  
\begin{equation} 
m_D^s=n_D^s-\sum_{\Gamma _S(|\Gamma _S|\geq 2)}\sum_{S_z=-S}^Sf_D^s(\Gamma 
_S^{S_z})m_{\Gamma _S^{S_z}}\;.  \label{f14} 
\end{equation} 
However, the additional conditions (I) and (II) prevent us from the 
derivation of an analytical expression for $m_D^s$, since the parameters $%
m_{\Gamma _S^{S_z}}$ depend on $m_D^s$ via (\ref{f7}) and (\ref{f12}). 
Therefore, the relation (\ref{f14}) has to be implemented into our 
minimization algorithm with the help of appropriate Lagrange parameters.

\section{Diagrams and Vertices} 
\label{integr} 
\subsection{The vertices $\tilde{x}_{\frak{D}}^{\frak{D}^{\prime }}$ and $%
\tilde{\xi}_{\frak{D}}^{\frak{D}^{\prime }\frak{D}^{\prime \prime }}$} 
 
In Section (\ref{calcu}) the vertices $\tilde{x}_{\frak{D}}^{\frak{D}%
^{\prime }}$ (see eqs. (\ref{4.14a})) have been derived in terms of the 
coefficients $x_{II^{\prime }}$, which occur in the expansion (\ref{1.8}). 
Now we will derive an explicit expression for these coefficients. The 
operator (\ref{4.18d}) may be written as  
\begin{eqnarray} 
m_{J^{\prime }}^{I_1,I_2} &=&\prod_{\sigma \in J^{\prime }}\left( n_\sigma 
^0+\hat{n}_\sigma ^{\text{HF}}\right) \prod_{\sigma \in \bar{J}^{\prime 
}\backslash (I_1\cup I_2)}\left( 1-n_\sigma ^0-\hat{n}_\sigma ^{\text{HF}%
}\right)  \label{in7a} \\ 
&=&\sum_{J(I_1,I_2\notin J)}\left[ \left( -1\right) ^{\left| J\cap \bar{J}%
^{\prime }\right| }\prod_{\sigma \in J^{\prime }\backprime J}n_\sigma 
^0\prod_{\sigma \in \bar{J}^{\prime }\backprime \{J\cup I_1\cup 
I_2\}}(1-n_\sigma ^0)\right] \hat{n}_J^{\text{HF}}\;.  \label{in7b} 
\end{eqnarray} 
Thus, $\hat{P}_{\text{G}}^2$ in (\ref{1.8}) becomes  
\begin{eqnarray} 
\hat{P}_{\text{G}}^2 &=&\sum_\Gamma \lambda _\Gamma ^2\sum_{I_1,I_2(I_1\cap 
I_2=\emptyset )}\sum_{J^{\prime }(I_1,I_2\notin J^{\prime })}T_{J^{\prime 
}\cup I_1,\Gamma }^{}T_{\Gamma ,J^{\prime }\cup I_2}^{+}f_{I_1}^{J^{\prime 
}}f_{I_2}^{J^{\prime }}\left( \prod_{\sigma \in I_1\cup I_2}\frac 
1{1-n_\sigma ^0}\right)  \label{in8} \\ 
&&\times \sum_{J(I_1,I_2\notin J)}\left[ \left( -1\right) ^{\left| J\cap  
\bar{J}^{\prime }\right| }\prod_{\sigma \in J^{\prime }\backprime J}n_\sigma 
^0\prod_{\sigma \in \bar{J}^{\prime }\backprime \{J\cup I_1\cup 
I_2\}}(1-n_\sigma ^0)\right] \hat{n}_{J\cup I_1,J\cup I_2}^{\text{HF}}\;.  
\nonumber 
\end{eqnarray} 
A comparison of the coefficients in (\ref{in8}) and (\ref{1.8}) gives  
\begin{eqnarray} 
x_{J\cup I_1,J\cup I_2} &=&\sum_{J^{\prime }(I_1,I_2\notin J^{\prime 
})}T_{J^{\prime }\cup I_1,\Gamma }^{}T_{\Gamma ,J^{\prime }\cup 
I_2}^{+}f_{I_1}^{J^{\prime }}f_{I_2}^{J^{\prime }}\left( \prod_{\sigma \in 
I_1\cup I_2}\frac 1{1-n_\sigma ^0}\right) \\ 
&&\times \left[ \left( -1\right) ^{\left| J\cap \bar{J}^{\prime }\right| 
}\prod_{\sigma \in J^{\prime }\backprime J}n_\sigma ^0\prod_{\sigma \in \bar{%
J}^{\prime }\backprime \{J\cup I_1\cup I_2\}}(1-n_\sigma ^0)\right] \;.  
\nonumber 
\end{eqnarray} 
In Appendix (\ref{interac}) we need an expression for vertices  
\begin{equation} 
\tilde{\xi}_{\frak{D}}^{\frak{D}^{\prime }\frak{D}^{\prime \prime }}=\tilde{%
\xi}_{(\sigma _1\sigma _2)}^{(\sigma _3\sigma _4)(\sigma _5\sigma _6)} 
\label{4.14aa} 
\end{equation} 
with three incoming and outgoing lines, which stem from a term like  
\begin{equation} 
y_{l;(\sigma _1,\sigma _3,\sigma _5)(\sigma _2,\sigma _4,\sigma _6)}\left\{ 
...\left( \hat{c}_{l,\sigma _1}^{+}\hat{c}_{l,\sigma _2}^{}\right) \left(  
\hat{c}_{l,\sigma _3}^{+}\hat{c}_{l,\sigma _4}^{}\right) \left( \hat{c}%
_{l,\sigma _5}^{+}\hat{c}_{l,\sigma _6}^{}\right) ...\right\} _{\Phi 
_0}^cf_{\sigma _1}^{\sigma _3}\,f_{\sigma _5}^{\sigma _1\cup \sigma 
_3}\,f_{\sigma _2}^{\sigma _4}\,f_{\sigma _6}^{\sigma _2\cup \sigma _4}\;. 
\label{4.14b} 
\end{equation} 
Hence, these vertices are given as  
\begin{equation} 
\tilde{\xi}_{(\sigma _1\sigma _2)}^{(\sigma _3\sigma _4)(\sigma _5\sigma 
_6)}=f_{\sigma _1}^{\sigma _3}\,f_{\sigma _5}^{\sigma _1\cup \sigma 
_3}\,f_{\sigma _2}^{\sigma _4}\,f_{\sigma _6}^{\sigma _2\cup \sigma 
_4}x_{(\sigma _1,\sigma _3,\sigma _5)(\sigma _2,\sigma _4,\sigma _6)}\;. 
\label{4.14c} 
\end{equation} 
 
\subsection{Diagrams} 
 
In Appendix (\ref{interac}) and (\ref{kinetic}) we need some diagrams, which 
can be evaluated in momentum-space as follows: 
 
\begin{enumerate} 
\item  ${}$  
\begin{mathletters} 
\begin{eqnarray} 
W_{(\sigma _3\sigma _4)(\sigma _5\sigma _6)}^{(\sigma _1\sigma _2)}(\vec{q}%
\,)\equiv \sum_{i\neq j\neq l}e^{i\vec{q}(\vec{R}_j-\vec{R}%
_l)}P_{i,j}^{\sigma _1\sigma _4}P_{j,l}^{\sigma _3\sigma _6}P_{l,i}^{\sigma 
_5\sigma _2} \\ 
=\sum_{\vec{k}}n_{\vec{k}}^{\sigma _1\sigma _4}n_{\vec{k}+\vec{q}}^{\sigma 
_3\sigma _6}n_{\vec{k}}^{\sigma _5\sigma _2}-\delta _{\sigma _3}^{\sigma 
_6}n_{\sigma _3}^0\tilde{P}_{(\sigma _5\sigma _4)}^{(\sigma _1\sigma _2)}(%
\vec{0}\,)-\delta _{\sigma _2}^{\sigma _5}n_{\sigma _2}^0\tilde{P}_{(\sigma 
_3\sigma _4)}^{(\sigma _1\sigma _6)}(\vec{q}\,)  \nonumber \\ 
-\delta _{\sigma _1}^{\sigma _4}n_{\sigma _1}^0\tilde{P}_{(\sigma _5\sigma 
_6)}^{(\sigma _3\sigma _2)}(\vec{q}\,)+2\delta _{\sigma _1}^{\sigma 
_4}\delta _{\sigma _3}^{\sigma _6}\delta _{\sigma _2}^{\sigma _5}n_{\sigma 
_1}^0n_{\sigma _2}^0n_{\sigma _3}^0 
\end{eqnarray} 
 
\item  ${}$  
\end{mathletters} 
\begin{equation} 
E_{(\sigma _1\sigma _2)}^{(\sigma _3\sigma _4)}(\vec{q}\,)\equiv 
\sum_{i(\neq )j}e^{i\vec{q}(\vec{R}_i-\vec{R}_j)}t_{i,j}^{\sigma _3\sigma 
_4}P_{i,j}^{\sigma _1\sigma _2}=\sum_{\vec{k}}\varepsilon _{\vec{k}}^{\sigma 
_3\sigma _4}n_{\vec{k}+\vec{q}}^{\sigma _1\sigma _2}\;, 
\end{equation} 
with  
\begin{equation} 
\varepsilon _{\vec{k}}^{\sigma _3\sigma _4}\equiv \sum_{i\neq j}e^{i\vec{k}(%
\vec{R}_i-\vec{R}_j)}t_{i,j}^{\sigma _3\sigma _4} 
\end{equation} 
 
\item  ${}$  
\begin{mathletters} 
\begin{eqnarray} 
V_{(\sigma _3\sigma _4)(\sigma _5\sigma _6)}^{(\sigma _1\sigma _2)}(\vec{q}%
\,)\equiv -\sum_{i\neq j\neq l}e^{i\vec{q}(\vec{R}_i-\vec{R}%
_l)}t_{i,j}^{\sigma _5\sigma _6}P_{i,l}^{\sigma _4\sigma _2}P_{l,j}^{\sigma 
_1\sigma _3} \\ 
=-\sum_{\vec{k}}\varepsilon _{\vec{k}}^{\sigma _5\sigma _6}n_{\vec{k}+\vec{q}%
}^{\sigma _4\sigma _2}n_{\vec{k}}^{\sigma _1\sigma _3}+\delta _{\sigma 
_2}^{\sigma _4}n_{\sigma _2}^0E_{(\sigma _1\sigma _3)}^{(\sigma _5\sigma 
_6)}(\vec{0}\,)+\delta _{\sigma _1}^{\sigma _3}n_{\sigma _1}^0E_{(\sigma 
_4\sigma _2)}^{(\sigma _5\sigma _6)}(\vec{q}\,) 
\end{eqnarray} 
 
and  
\end{mathletters} 
\begin{mathletters} 
\begin{eqnarray} 
\bar{V}_{(\sigma _3\sigma _4)(\sigma _5\sigma _6)}^{(\sigma _1\sigma _2)}(%
\vec{q}\,)\equiv -\sum_{i\neq j\neq l}e^{i\vec{q}(\vec{R}_l-\vec{R}%
_j)}t_{i,j}^{\sigma _5\sigma _6}P_{i,l}^{\sigma _4\sigma _2}P_{l,j}^{\sigma 
_1\sigma _3} \\ 
=-\sum_{\vec{k}}\varepsilon _{\vec{k}}^{\sigma _5\sigma _6}n_{\vec{k}%
}^{\sigma _4\sigma _2}n_{\vec{k}+\vec{q}}^{\sigma _1\sigma _3}+\delta 
_{\sigma _2}^{\sigma _4}n_{\sigma _2}^0E_{(\sigma _1\sigma _3)}^{(\sigma 
_5\sigma _6)}(\vec{q}\,)+\delta _{\sigma _1}^{\sigma _3}n_{\sigma 
_1}^0E_{(\sigma _4\sigma _2)}^{(\sigma _5\sigma _6)}(\vec{0}\,) 
\end{eqnarray} 
 
\item  ${}$  
\end{mathletters} 
\begin{mathletters} 
\begin{eqnarray} 
U_{(\sigma _3\sigma _4)(\sigma _5\sigma _6)}^{(\sigma _1\sigma _2)}(\vec{q}%
\,)\equiv \sum_{i\neq j\neq l\neq m}e^{i\vec{q}(\vec{R}_m-\vec{R}%
_l)}t_{i,j}^{\sigma _1\sigma _2}P_{i,m}^{\sigma _1\sigma _4}P_{m,l}^{\sigma 
_3\sigma _6}P_{l,j}^{\sigma _5\sigma _2} \\ 
=\sum_{\vec{k}}\varepsilon _{\vec{k}}^{\sigma _1\sigma _2}n_{\vec{k}%
}^{\sigma _1\sigma _4}n_{\vec{k}+\vec{q}}^{\sigma _3\sigma _6}n_{\vec{k}%
}^{\sigma _5\sigma _2}-\delta _{\sigma _3}^{\sigma _6}n_{\sigma _3}^0\bar{V}%
_{(\sigma _1\sigma _2)(\sigma _1\sigma _2)}^{(\sigma _5\sigma _4)}(\vec{0}%
\,)-\delta _{\sigma _1}^{\sigma _4}n_{\sigma _1}^0V_{(\sigma _3\sigma 
_2)(\sigma _1\sigma _2)}^{(\sigma _5\sigma _6)}(\vec{q}\,)+ \\ 
-\delta _{\sigma _2}^{\sigma _5}n_{\sigma _2}^0\bar{V}_{(\sigma _1\sigma 
_6)(\sigma _1\sigma _2)}^{(\sigma _3\sigma _4)}(\vec{q}\,)-\delta _{\sigma 
_1}^{\sigma _4}\delta _{\sigma _2}^{\sigma _5}n_{\sigma _1}^0n_{\sigma 
_2}^0E_{(\sigma _3\sigma _6)}^{(\sigma _1\sigma _2)}(\vec{q}\,)-\delta 
_{\sigma _1}^{\sigma _4}\delta _{\sigma _3}^{\sigma _6}n_{\sigma 
_1}^0n_{\sigma _3}^0E_{(\sigma _5\sigma _2)}^{(\sigma _1\sigma _2)}(\vec{0}%
\,)+  \nonumber \\ 
-\delta _{\sigma _2}^{\sigma _5}\delta _{\sigma _3}^{\sigma _6}n_{\sigma 
_1}^0n_{\sigma _3}^0E_{(\sigma _1\sigma _4)}^{(\sigma _1\sigma _2)}(\vec{0}%
\,)-\tilde{P}_{(\sigma _3\sigma _4)}^{(\sigma _1\sigma _6)}(\vec{q}%
\,)E_{(\sigma _5\sigma _2)}^{(\sigma _1\sigma _2)}(\vec{q}\,)-\tilde{P}%
_{(\sigma _3\sigma _2)}^{(\sigma _5\sigma _6)}(\vec{q}\,)E_{(\sigma _1\sigma 
_4)}^{(\sigma _1\sigma _2)}(\vec{q}\,)  \nonumber 
\end{eqnarray} 
\end{mathletters} 
\end{enumerate}

\section{Evaluation of the atomic interaction} 
\label{interac} 
According to eq. (\ref{4.2a}) we have to analyze the expectation values  
\begin{mathletters} 
\label{4.24} 
\begin{eqnarray} 
\frac 1{N_{\vec{q}}}\sum_k\left\langle \Psi _{\vec{q}}^{\text{G}}\left| \hat{%
m}_{k;\Gamma }\right| \Psi _{\vec{q}}^{\text{G}}\right\rangle  &=&\frac 1{N_{%
\vec{q}}}\sum_{i,j,k}e^{i\vec{q}(\vec{R}_i-\vec{R}_j)}\left\langle \Psi _{%
\text{G}}\left| \hat{S}_i^{+}\hat{m}_{k;\Gamma }\hat{S}_j^{-}\right| \Psi _{%
\text{G}}\right\rangle   \label{4.24a} \\ 
&=&\frac 1{N_{\vec{q}}}\sum_k\Bigl[\sum_{i,j(\neq k)}e^{i\vec{q}(\vec{R}_i-%
\vec{R}_j)}\left\langle \Psi _{\text{G}}\left| \hat{S}_i^{+}\hat{S}_j^{-}%
\hat{m}_{k;\Gamma }\right| \Psi _{\text{G}}\right\rangle   \label{4.24b} \\ 
&&+\sum_{j(\neq k)}\left( e^{i\vec{q}(\vec{R}_k-\vec{R}_j)}\left\langle \Psi 
_{\text{G}}\left| \hat{S}_k^{+}\hat{m}_{k;\Gamma }\hat{S}_j^{-}\right| \Psi 
_{\text{G}}\right\rangle +c.c.\right)   \label{4.24c} \\ 
&&+\left\langle \Psi _{\text{G}}\left| \hat{S}_k^{+}\hat{m}_{k;\Gamma }\hat{S%
}_k^{-}\right| \Psi _{\text{G}}\right\rangle \Bigr]  \label{4.24d} 
\end{eqnarray} 
This evaluation will be done separately for the three terms (\ref{4.24b})-(%
\ref{4.24d}). 
 
In the first term (\ref{4.24b}), we have to distinguish connected and 
unconnected diagrams. Here, the term ``connected'' means, that the lattice 
site $k$ is connected to one of the lattice sites $i,j$. This condition 
necessarily requires that $k$ is connected to both lattice sites $i$ and $j$%
, because $\hat{S}_i^{+}$ or $\hat{S}_j^{-}$ generate a spin-flip. Such a 
process cannot be compensated in a diagram with only one of these operators, 
neither by the external vertex-operator $\hat{m}_{k;\Gamma }$ nor by one of 
the internal vertex-operators $\hat{n}_{l;I,I^{\prime }}^{\text{HF}}$. 
 
The unconnected terms can be written as  
\end{mathletters} 
\begin{eqnarray} 
(\ref{4.24b})^{\text{uc}} &=&\frac 1{N_{\vec{q}}}\sum_k\frac{\left\langle 
\Psi _{\text{G}}\left| \hat{m}_{k;\Gamma }\right| \Psi _{\text{G}%
}\right\rangle }{N_{\text{G}}}  \label{4.25} \\ 
&&\times \Bigl[\sum_{i\neq j(\neq k)}e^{i\vec{q}(\vec{R}_i-\vec{R}%
_j)}\prod_{l(\neq i,j,k)}\left\langle \left( \hat{P}_{i;\text{G}}\hat{S}%
_i^{+}\hat{P}_{i;\text{G}}\right) \left( \hat{P}_{j;\text{G}}\hat{S}_j^{-}%
\hat{P}_{j;\text{G}}\right) \,\hat{P}_{l;\text{G}}^2\right\rangle _{\Phi _0} 
\nonumber \\ 
&&+\sum_i\prod_{l(\neq i,k)}\left\langle \left( \hat{P}_{i;\text{G}}\hat{S}%
_i^{+}\hat{S}_i^{-}\hat{P}_{i;\text{G}}\right) \,\hat{P}_{l;\text{G}%
}^2\right\rangle _{\Phi _0}\Bigr]\;.  \nonumber 
\end{eqnarray} 
When we ignore the restriction $i,j,l\neq k$, the sum over $i$ and $j$ gives 
just $N_{\vec{q}}$. Thus, we find the correct result for $(\ref{4.24b})^{%
\text{uc}}$ by substracting all diagrams with $i=k$ or $j=k$ for an external 
vertex, or $l=k$ for one of the internal vertices. These contributions are 
given as  
\[ 
\begin{tabular}{ccl} 
$i=j=k$ & $:$ & $N_{\text{G}}\sum_{\Gamma ^{\prime }}S_{-}(\Gamma 
^{\prime })m_{\Gamma ^{\prime }}$ \\  
$i=k\text{ or }j=k$ & $:$ & $2N_{\text{G}}\vec{S}^{+}\cdot \tilde{\Omega}(%
\vec{q}\,)\cdot \vec{S}^{-}$ \\  
$l=k$ & $:$ & $N_{\text{G}}\vec{S}^{+}\cdot \tilde{\Omega}(\vec{q}\,)\cdot \,%
\tilde{x}\cdot \tilde{\Omega}(\vec{q}\,)\cdot \vec{S}^{-}\;.$%
\end{tabular} 
\] 
Altogether, we obtain the following expression for the contribution of the 
unconnected diagrams:  
\begin{mathletters} 
\label{4.24aa} 
\begin{eqnarray} 
(\ref{4.24b})^{\text{uc}} &=&\left[ Lm_\Gamma -m_\Gamma \frac{LN_{\text{G}}}{%
N_{\vec{q}}}\Bigl(2\vec{S}^{+}\cdot \tilde{\Omega}(\vec{q}\,)\cdot \vec{S}%
^{-}+\sum_{\Gamma ^{\prime }}S_{-}(\Gamma ^{\prime })m_{\Gamma ^{\prime 
}}+\vec{S}^{+}\cdot \tilde{\Omega}(\vec{q}\,)\cdot \tilde{x}\cdot \tilde{%
\Omega}(\vec{q}\,)\cdot \vec{S}^{-}\Bigr)\right]  \\ 
&\equiv &Lm_\Gamma +m_\Gamma ^1(\vec{q})\;, 
\end{eqnarray} 
where $\frac{LN_{\text{G}}}{N_{\vec{q}}}$ is given in (\ref{4.23d}). Note 
that only the second term $m_\Gamma ^1(\vec{q})$ is relevant for our 
spin-wave-dispersion $E_{\vec{q}}^{var}$, since the first term $Lm_\Gamma $ 
is canceled by the respective ground-state contribution in eq. (\ref{3.7}). 
 
For the connected diagrams in (\ref{4.24b}) we may distinguish between those 
diagrams with two or four lines, which enter or leave the external vertex. 
The vertex-factors with two lines can be evaluated from (\ref{4.20}), 
whereas the respective factor with four lines stems from the expansion  
\end{mathletters} 
\begin{eqnarray} 
\left\{ ...\hat{m}_{I_1,I_2}...\right\} _{\Phi _0}^C &\rightarrow 
&\sum_{\sigma _1,\sigma _2,\sigma _3,\sigma _4}\tilde{\frak{V}}%
_{_{I_1,I_2}}^{(\sigma _1\sigma _2)(\sigma _3\sigma _4)}\left\{ ...\left(  
\hat{c}_{\sigma _1}^{+}\hat{c}_{\sigma _2}\right) \left( \hat{c}_{\sigma 
_3}^{+}\hat{c}_{\sigma _4}\right) ...\right\} _{\Phi _0}^c  \label{4.23a} \\ 
\tilde{\frak{V}}_{_{I_1,I_2}}^{(\sigma _1\sigma _2)(\sigma _3\sigma _4)} 
&=&\sum_{J(\sigma _1,\sigma _2,\sigma _3,\sigma _4\notin I)}f_{\sigma 
_1}^J\,f_{\sigma _2}^J\,f_{\sigma _3}^J\,f_{\sigma _4}^Jm_J^0\Bigl%
(\prod_{\sigma \in (\sigma _1,\sigma _2,\sigma _3,\sigma _4)}\frac 
1{(1-n_\sigma ^0)}\Bigr) \\ 
&&\times \left[ f_{\sigma _1}^{\sigma _3}\,f_{\sigma _2}^{\sigma _4}\delta 
_{J\cup \left( \sigma _1,\sigma _3\right) }^{I_1}\delta _{J\cup \left( 
\sigma _2,\sigma _4\right) }^{I_2}-\delta _{\sigma _1}^{\sigma _2}\delta 
_{J\cup \sigma _3}^{I_1}\delta _{J\cup \sigma _4}^{I_2}-\delta _{\sigma 
_3}^{\sigma _4}\delta _{J\cup \sigma _1}^{I_1}\delta _{J\cup \sigma 
_2}^{I_2}+\right.   \nonumber \\ 
&&\left. -\delta _{\sigma _1}^{\sigma _4}\delta _{J\cup \sigma 
_1}^{I_1}\delta _{J\cup \sigma _2}^{I_2}-\delta _{\sigma _2}^{\sigma 
_3}\delta _{J\cup \sigma _1}^{I_1}\delta _{J\cup \sigma _4}^{I_2}+(\delta 
_{\sigma _1}^{\sigma _2}\delta _{\sigma _3}^{\sigma _4}-\delta _{\sigma 
_1}^{\sigma _4}\delta _{\sigma _2}^{\sigma _3})\delta _J^{I_1}\delta 
_J^{I_2}\right] \;,  \nonumber 
\end{eqnarray} 
which is also derived for $\left| I_1\right| =\left| I_2\right| $. This 
expression together with eq. (\ref{4.18b}) leads to the following external 
vertex $\stackrel{+}{T}{\!}_{\frak{D}}^{\frak{D}^{\prime }}$ for the 
operator $\hat{S}_i^{+}$%
\begin{equation} 
\stackrel{+}{T}{\!}_{(\sigma _1\sigma _2)}^{(\sigma _3\sigma 
_4)}=\sum_\Gamma \lambda _{\Gamma _{+}}\lambda _\Gamma \sqrt{S_{+}(\Gamma 
)}\sum_{I_1,I_2}T_{I_1,\Gamma _{+}}^{}T_{\Gamma ,I_2}^{+}{\frak{\tilde{V}}}%
_{_{I_1,I_2}}^{(\sigma _1\sigma _2)(\sigma _3\sigma _4)}\;,  \label{4.23b} 
\end{equation} 
whereas the corresponding factor $\stackrel{-}{T}{\!}_{\frak{D}}^{\frak{D}%
^{\prime }}$ for the operator $\hat{S}_j^{-}$ is given as  
\begin{equation} 
\stackrel{-}{T}{\!}_{(\sigma _1\sigma _2)}^{(\sigma _3\sigma _4)}=\left(  
\stackrel{+}{T}{\!}_{(\sigma _2\sigma _1)}^{(\sigma _4\sigma _3)}\right) 
^{*}\;.  \label{4.23c} 
\end{equation} 
Using eqs. (\ref{4.20}), (\ref{4.23a}), and  
\begin{equation} 
\hat{P}_{\text{G}}\hat{m}_\Gamma \hat{P}_{\text{G}}=\sum_{I_1,I_2}\lambda 
_\Gamma ^2T_{\Gamma ,I_1}^{}T_{I_2,\Gamma }^{+}\hat{m}_{I_1,I_2} 
\end{equation} 
we obtain the following expression for the vertices of the external-operator 
in (\ref{4.24b}) with one or two incoming and outgoing lines,  
\begin{mathletters} 
\label{4.24bb} 
\begin{eqnarray} 
M_{(\sigma _1\sigma _2)}(\Gamma ) &=&\sum_{I_1,I_2}\lambda _\Gamma 
^2T_{\Gamma ,I_1}^{}T_{I_2,\Gamma }^{+}{\frak{V}}_{_{I_1,I_2}}^{(\sigma 
_1\sigma _2)}\;, \\ 
\tilde{M}_{(\sigma _1\sigma _2)}^{(\sigma _3\sigma _4)}(\Gamma ) 
&=&\sum_{I_1,I_2}\lambda _\Gamma ^2T_{\Gamma ,I_1}^{}T_{I_2,\Gamma }^{+}%
\tilde{\frak{V}}_{_{I_1,I_2}}^{(\sigma _1\sigma _2)(\sigma _3\sigma _4)}\;. 
\end{eqnarray} 
When we define the vector $\vec{M}(\Gamma )$ of components $M_{\frak{D}%
}(\Gamma )$, we may write the connected terms (\ref{4.24b}) with $i\neq j$ 
as  
\end{mathletters} 
\begin{mathletters} 
\label{4.24cc} 
\begin{eqnarray} 
m_\Gamma ^2(\vec{q}\,) &=&\frac{LN_{\text{G}}}{N_{\vec{q}}}\sum_{\frak{D}_1,%
\frak{D}_2,\frak{D}_3}\left[ \vec{M}(\Gamma )\cdot \left( \hat{1}-\tilde{%
\Omega}(\vec{0}\,)\cdot \tilde{x}\right) \right] _{\frak{D}_1}\left( W_{%
\frak{D}_2,\frak{D}_3}^{\frak{D}_1}(\vec{q}\,)+W_{\frak{D}_3,\frak{D}_2}^{%
\frak{D}_1}(\vec{q}\,)\right)   \nonumber \\ 
&&\times \left[ \left( \hat{1}+\tilde{x}\cdot \tilde{\Omega}(\vec{q}%
\,)\right) \cdot \vec{S}^{+}\right] _{\frak{D}_2}\left[ \left( \hat{1}+%
\tilde{x}\cdot \tilde{\Omega}(\vec{q}\,)\right) \cdot \vec{S}^{-}\right] _{%
\frak{D}_3} \\ 
m_\Gamma ^3(\vec{q}\,) &=&\frac{LN_{\text{G}}}{N_{\vec{q}}}\sum_{\frak{D}_1,%
\frak{D}_2,\frak{D}_3}\left[ \vec{M}(\Gamma )\cdot \tilde{\Omega}(\vec{0}%
\,)\right] _{\frak{D}_1}\tilde{\xi}_{\frak{D}_1}^{\frak{D}_2,\frak{D}%
_3}\left[ \tilde{\Omega}(\vec{q}\,)\cdot \vec{S}^{+}\right] _{\frak{D}%
_2}\left[ \tilde{\Omega}(\vec{q}\,)\cdot \vec{S}^{-}\right] _{\frak{D}_3} \\ 
m_\Gamma ^4(\vec{q}\,) &=&\frac{LN_{\text{G}}}{N_{\vec{q}}}\vec{M}(\Gamma 
)\cdot \tilde{\Omega}(\vec{0}\,)\cdot \left( \stackrel{+}{T}\cdot \tilde{%
\Omega}(\vec{q}\,)\cdot \vec{S}^{-}+\stackrel{-}{T}\cdot \tilde{\Omega}(\vec{%
q}\,)\cdot \vec{S}^{+}\right)  \\ 
m_\Gamma ^5(\vec{q}\,) &=&\frac{LN_{\text{G}}}{N_{\vec{q}}}\vec{S}^{+}\cdot  
\tilde{\Omega}(\vec{q}\,)\cdot \tilde{M}(\Gamma )\cdot \tilde{\Omega}(\vec{q}%
\,)\cdot \vec{S}^{-} 
\end{eqnarray} 
The tensors $W_{\frak{D}_2,\frak{D}_3}^{\frak{D}_1}(\vec{q}\,)$, $\tilde{\xi}%
_{\frak{D}_1}^{\frak{D}_2,\frak{D}_3}$ are defined in Appendix (\ref{integr}%
). In fig. (\ref{fig4}) all diagrams, which belong to the atomic 
interactions are presented. 
 
The connected diagrams in (\ref{4.24b}) with $i=j$ are determined by the 
external vertex-operator  
\end{mathletters} 
\begin{equation} 
\hat{P}_{\text{G}}\hat{S}^{+}\hat{S}^{-}\hat{P}_{\text{G}}=\sum_{\Gamma 
^{\prime }}\lambda _{\Gamma ^{\prime }}^{2}S_{-}(\Gamma ^{\prime })\hat{m}%
_{\Gamma ^{\prime }}\;.  \label{4.30} 
\end{equation} 
This expression leads to the contribution  
\begin{equation} 
m_{\Gamma }^{6}(\vec{q}\,)=\frac{LN_{\text{G}}}{N_{\vec{q}}}\sum_{\Gamma 
^{\prime }}S_{-}(\Gamma ^{\prime })\left[ \vec{M}(\Gamma ^{\prime })\cdot  
\tilde{\Omega}(\vec{0}\,)\cdot \vec{M}(\Gamma )\right] =m_{\Gamma }^{6}(\vec{%
0}\,)\;. 
\end{equation} 
 
For the evaluation of (\ref{4.24c}), we need the external vertex for the 
operator $\hat{P}_{\text{G}}\hat{S}^{+}\hat{m}_\Gamma \hat{P}_{\text{G}}$, 
which is the same as the respective term in (\ref{4.21}) for fixed $\Gamma $%
,  
\begin{equation} 
A_{\left( \sigma _1\sigma _2\right) }^{+}\equiv \lambda _{\Gamma 
_{+}}\lambda _\Gamma \sqrt{S_{+}(\Gamma )}\sum_{I_1,I_2}T_{I_1,\Gamma 
_{+}}^{}T_{\Gamma ,I_2}^{+}{\frak{V}}_{_{I_1,I_2}}^{(\sigma _1\sigma _2)}\;. 
\end{equation} 
The only connected diagram in (\ref{4.24c}) is therefore given as  
\begin{equation} 
m_\Gamma ^7(\vec{q}\,)=\frac{LN_{\text{G}}}{N_{\vec{q}}}\left( \vec{A}%
^{+}\cdot \tilde{\Omega}(\vec{q}\,)\cdot \vec{S}^{-}+c.c.\right) \;. 
\end{equation} 
Finally, we determine the contribution (\ref{4.24d}) as  
\begin{equation} 
m_\Gamma ^8(\vec{q}\,)=\frac{LN_{\text{G}}}{N_{\vec{q}}}S_{+}(\Gamma 
)m_{\Gamma _{+}}=m_\Gamma ^8(\vec{0}\,)\;. 
\end{equation} 
Here, we used the relation  
\begin{equation} 
\hat{S}^{+}\hat{m}_\Gamma \hat{S}^{-}=S_{+}(\Gamma )\hat{m}_{\Gamma 
_{+}}\;. 
\end{equation} 
To summarize, the expectation value (\ref{4.24d}) for the atomic energy is 
given as  
\begin{equation} 
\frac{\left\langle \Psi _{\vec{q}}^{\text{G}}\left| \hat{H}_{\text{at}%
}\right| \Psi _{\vec{q}}^{\text{G}}\right\rangle }{N_{\vec{q}}}=L\sum_\Gamma 
E_\Gamma m_\Gamma +\sum_\Gamma E_\Gamma \sum_{c=1}^8m_\Gamma ^c(\vec{q}\,)\;. 
\end{equation}

\section{Evaluation of the one-particle energy} 
\label{kinetic} 
 
For the diagrammatic evaluation of the one-particle-energy we write the 
expectation value (\ref{4.2b}) as  
\begin{mathletters} 
\label{5.1} 
\begin{eqnarray} 
&&\frac{\left\langle \Psi _{\vec{q}}^{\text{G}}\left| \hat{H}_1\right| \Psi 
_{\vec{q}}^{\text{G}}\right\rangle }{N_{\vec{q}}}=\frac 1{N_{\vec{q}%
}}\sum_{k(\neq )l}\sum_{\sigma _k,\sigma _l}t_{k,l}^{\sigma _k,\sigma _l}%
\Bigl\{\sum_{i,j(\neq k,l)}e^{i\vec{q}(\vec{R}_i-\vec{R}_j)}\left\langle  
\hat{S}_i^{+}\hat{c}_{k;\sigma _k}^{+}\hat{c}_{l;\sigma _l}^{}\hat{S}%
_j^{-}\right\rangle _{\Psi _{\text{G}}}+  \label{5.1a} \\ 
&&+\sum_{i(\neq k,l)}\left( e^{i\vec{q}(\vec{R}_i-\vec{R}_l)}\left\langle  
\hat{S}_i^{+}\hat{c}_{k;\sigma _k}^{+}\hat{c}_{l;\sigma _l}^{}\hat{S}%
_l^{-}\right\rangle _{\Psi _{\text{G}}}+e^{i\vec{q}(\vec{R}_i-\vec{R}%
_k)}\left\langle \hat{S}_i^{+}\hat{c}_{k;\sigma _k}^{+}\hat{c}_{l;\sigma 
_l}^{}\hat{S}_k^{-}\right\rangle _{\Psi _{\text{G}}}\right) +c.c 
\label{5.1b} \\ 
&&+\left( e^{i\vec{q}(\vec{R}_k-\vec{R}_l)}\left\langle \hat{S}_k^{+}\hat{c}%
_{k;\sigma _k}^{+}\hat{c}_{l;\sigma _l}^{}\hat{S}_l^{-}\right\rangle _{\Psi 
_{\text{G}}}+e^{i\vec{q}(\vec{R}_l-\vec{R}_k)}\left\langle \hat{c}_{k;\sigma 
_k}^{+}\hat{S}_k^{-}\hat{S}_l^{+}\hat{c}_{l;\sigma _l}^{}\right\rangle 
_{\Psi _{\text{G}}}\right) +c.c  \label{5.1c} \\ 
&&+\left\langle \hat{S}_k^{+}\hat{c}_{k;\sigma _k}^{+}\hat{S}_k^{-}\hat{c}%
_{l;\sigma _l}^{}\right\rangle _{\Psi _{\text{G}}}+c.c.\Bigr\}  \label{5.1d} 
\end{eqnarray} 
Here, we already used the fact that the tight-binding parameters do not 
contain any local terms. In (\ref{5.1a}) we have to distinguish connected 
and unconnected diagrams. The unconnected contributions may be written as  
\end{mathletters} 
\begin{eqnarray} 
(\ref{5.1a})^{\text{uc}} &=&\frac 1{N_{\vec{q}}}\sum_{k\neq l}\sum_{\sigma 
_k,\sigma _l}t_{k,l}^{\sigma _k,\sigma _l}\left\langle \hat{S}_i^{+}\hat{c}%
_{k;\sigma _k}^{+}\hat{c}_{l;\sigma _l}^{}\hat{S}_j^{-}\right\rangle _{\Psi 
_{\text{G}}}  \label{5.2} \\ 
&&\times \Bigl[\sum_{i\neq j(\neq k,l)}e^{i\vec{q}(\vec{R}_i-\vec{R}%
_j)}\prod_{m(\neq i,j,k,l)}\left\langle \left( \hat{P}_{i;\text{G}}\hat{S}%
_i^{+}\hat{P}_{i;\text{G}}\right) \left( \hat{P}_{j;\text{G}}\hat{S}_j^{-}%
\hat{P}_{j;\text{G}}\right) \,\hat{P}_{m;\text{G}}^2\right\rangle _{\Phi _0} 
\nonumber \\ 
&&+\sum_i\prod_{m(\neq i,k,l)}\left\langle \left( \hat{P}_{i;\text{G}}\hat{S}%
_i^{+}\hat{S}_i^{-}\hat{P}_{i;\text{G}}\right) \,\hat{P}_{m;\text{G}%
}^2\right\rangle _{\Phi _0}\Bigr]\;.  \nonumber 
\end{eqnarray} 
Eq. (\ref{5.2}) can be evaluated, using the same arguments as discussed in 
connection with eq. (\ref{4.25}). This evaluation leads to  
\begin{mathletters} 
\label{5.1bb} 
\begin{eqnarray} 
(\ref{5.1a})^{\text{uc}} &=&E_{\text{kin}}-\varepsilon ^1(\vec{q}\,) 
\label{5.2b} \\ 
\varepsilon ^1(\vec{q}\,) &=&\frac{2E_{\text{kin}}}L\frac{LN_{\text{G}}}{N_{%
\vec{q}}}\left( 2\vec{S}^{+}\cdot \tilde{\Omega}(\vec{q}\,)\cdot \vec{S}%
^{-}+\sum_{\Gamma ^{\prime }}S_{-}(\Gamma ^{\prime })m_{\Gamma ^{\prime 
}}+\vec{S}^{+}\cdot \tilde{\Omega}(\vec{q}\,)\cdot \tilde{x}\cdot \tilde{%
\Omega}(\vec{q}\,)\cdot \vec{S}^{-}\right)  
\end{eqnarray} 
where  
\end{mathletters} 
\begin{equation} 
E_{\text{kin}}=\frac 1{N_{\text{G}}}\sum_{k\neq l}\sum_{\sigma _k,\sigma 
_l}t_{k,l}^{\sigma _k,\sigma _l}\left\langle \hat{c}_{k;\sigma _k}^{+}\hat{c}%
_{l;\sigma _l}^{}\right\rangle _{\Psi _{\text{G}}} 
\end{equation} 
 
Before we start to discuss the connected diagrams, we should first consider 
the general structure of external vertices in (\ref{5.1a}) at lattice sites $%
k$ and $l$, where electrons are created or annihilated. For example, the 
vertex-function of the operator $\hat{c}_\sigma ^{+}$ leads to  
\begin{equation} 
\left\{ ...\hat{P}_{\text{G}}\hat{c}_\sigma ^{+}\hat{P}_{\text{G}%
}...\right\} _{\Phi _0}^C=\sum_{\Gamma ,\Gamma ^{\prime }}\lambda _\Gamma 
\lambda _{\Gamma ^{\prime }}\sum_{I(\sigma \notin I)}f_\sigma ^IT_{\Gamma 
,I}^{+}T_{I\cup \sigma ,\Gamma ^{\prime }}^{}\sum_{I_1,I_2}T_{I_1,\Gamma 
}^{}T_{\Gamma ^{\prime },I_4}^{+}\left\{ ...\hat{m}_{I_1,I_2}...\right\} 
_{\Phi _0}^C  \label{5.3} 
\end{equation} 
Hence, our general problem is the calculation of vertex-contributions for 
operators $\hat{m}_{I_1,I_2}$ with $\left| I_1\right| -\left| I_2\right| =1$%
. The first case is a vertex with only one incoming and no outgoing line,  
\begin{mathletters} 
\label{5.4} 
\begin{eqnarray} 
\left\{ ...\hat{m}_{I_1,I_2}...\right\} _{\Phi _0}^C &\rightarrow 
&\sum_{\sigma ^{\prime }}{{\frak{h}}}_{_{I_1,I_2}}(\sigma ^{\prime })\left\{ 
...\hat{c}_{\sigma _1}^{+}...\right\} _{\Phi _0}^c  \label{5.4a} \\ 
{\frak{h}}_{I_1,I_2}(\sigma ^{\prime }) &=&\frac 1{1-n_{\sigma ^{\prime 
}}^0}\sum_{I^{\prime }(\sigma ^{\prime }\notin I^{\prime })}f_{\sigma 
^{\prime }}^{I^{\prime }}m_{I^{\prime }}^0\delta _{I_1}^{I^{\prime }\cup 
\sigma ^{\prime }}\delta _{I_2}^{I^{\prime }}\;.  \label{5.4b} 
\end{eqnarray} 
Note that eq. (\ref{5.3}) together with (\ref{qfac}) yields directly the $q$%
-factor (\ref{qfac}), which occurs as the renormalization factor for 
hopping-processes. 
 
Furthermore we have the cases with two (three) incoming and one (two) 
outgoing lines. This processes lead to the vertices  
\end{mathletters} 
\begin{mathletters} 
\label{5.4bb} 
\begin{eqnarray} 
\left\{ ...\hat{m}_{I_1,I_2}...\right\} _{\Phi _0}^C &\rightarrow 
&\sum_{\sigma ^{\prime }}\sum_{\sigma _1,\sigma _2}{\frak{H}}%
_{I_1,I_2}^{\left( \sigma _1,\sigma _2\right) }(\sigma ^{\prime })\left\{ 
...\left( \hat{c}_{\sigma ^{\prime }}^{+}\right) \left( \hat{c}_{\sigma 
_1}^{+}\hat{c}_{\sigma _2}^{}\right) ...\right\} _{\Phi _0}^c  \label{5.5} \\ 
{\frak{H}}_{I_1,I_2}^{\left( \sigma _1\sigma _2\right) }(\sigma ^{\prime }) 
&=&\sum_{I^{\prime }(\sigma ^{\prime },\sigma _1,\sigma _2\notin I^{\prime 
})}f_{\sigma ^{\prime }}^{I^{\prime }}f_{\sigma _1}^{I^{\prime }}\,f_{\sigma 
_2}^{I^{\prime }}m_{^{I^{\prime }}}^0\Bigl(\prod_{\tilde{\sigma}\in (\sigma 
^{\prime },\sigma _1,\sigma _2)}\frac 1{(1-n_{\tilde{\sigma}}^0)}\Bigr) 
\label{5.5b} \\ 
&&\times \left[ f_{\sigma ^{\prime }}^{\sigma _1}\delta _{I_1,}^{I^{\prime 
}\cup \left( \sigma ^{\prime },\sigma _1\right) }\delta _{I_2}^{I^{\prime 
}\cup \sigma _2}-\delta _{\sigma _1}^{\sigma _2}\delta _{I_1}^{I^{\prime 
}\cup \sigma ^{\prime }}\delta _{I_2}^{I^{\prime }}+\delta _{\sigma ^{\prime 
}}^{\sigma _2}\delta _{I_1}^{I^{\prime }\cup \sigma _1}\delta 
_{I_2}^{I^{\prime }}\right] \;.  \nonumber 
\end{eqnarray} 
and  
\end{mathletters} 
\begin{mathletters} 
\label{5.4cc} 
\begin{eqnarray} 
&&\left\{ ...\hat{m}_{I_1,I_2}...\right\} _{\Phi _0}^C\rightarrow 
\sum_{\sigma ^{\prime }}\sum_{\sigma _1,\sigma _2}{\frak{\tilde{H}}}%
_{I_1,I_2}^{\left( \sigma _1\sigma _2\right) (\sigma _3\sigma _4)}(\sigma 
^{\prime })\left\{ ...\left( \hat{c}_{\sigma ^{\prime }}^{+}\right) \left(  
\hat{c}_{\sigma _1}^{+}\hat{c}_{\sigma _2}^{}\right) \left( \hat{c}_{\sigma 
_3}^{+}\hat{c}_{\sigma _4}^{}\right) ...\right\} _{\Phi _0}^c  \label{5.6} \\ 
&&{\frak{\tilde{H}}}_{I_1,I_2}^{\left( \sigma _1\sigma _2\right) (\sigma 
_3\sigma _4)}(\sigma ^{\prime })=\sum_{I(\sigma _1,\sigma _2,\sigma 
_3,\sigma _4,\sigma ^{\prime }\notin I)}f_{\sigma _1}^I\,f_{\sigma 
_2}^I\,f_{\sigma _3}^I\,f_{\sigma _4}^If_{\sigma ^{\prime }}^Im_I^0\Bigl%
(\prod_{\sigma \in (\sigma _1,\sigma _2,\sigma _3,\sigma _4,\sigma ^{\prime 
})}\frac 1{(1-n_\sigma ^0)}\Bigr)f_{\sigma ^{\prime }}^{\sigma _1}f_{\sigma 
^{\prime }}^{\sigma _3} \\ 
&&\times \Bigl[f_{\sigma _1}^{\sigma _3}\,f_{\sigma _2}^{\sigma _4}\delta 
_{I\cup \left( \sigma _1,\sigma _3,\sigma ^{\prime }\right) }^{I_1}\delta 
_{I\cup \left( \sigma _2,\sigma _4\right) ,}^{I_2}-f_{\sigma ^{\prime 
}}^{\sigma _1}\left( \delta _{\sigma _1}^{\sigma _2}\delta _{I\cup (\sigma 
_3,\sigma ^{\prime })}^{I_1}\delta _{I\cup \sigma _4}^{I_2}-\delta _{\sigma 
_1}^{\sigma _4}\delta _{I\cup (\sigma _3,\sigma ^{\prime })}^{I_1}\delta 
_{I\cup \sigma _2}^{I_2}\right)   \nonumber \\ 
&&+f_{\sigma ^{\prime }}^{\sigma _3}\left( \delta _{\sigma _2}^{\sigma 
_3}\delta _{I\cup (\sigma _1,\sigma ^{\prime })}^{I_1}\delta _{I\cup \sigma 
_4}^{I_2}-\delta _{\sigma _3}^{\sigma _4}\delta _{I\cup (\sigma _1,\sigma 
^{\prime })}^{I_1}\delta _{I\cup \sigma _2}^{I_2}\right) +f_{\sigma ^{\prime 
}}^{\sigma _1}f_{\sigma ^{\prime }}^{\sigma _3}(\delta _{\sigma _1}^{\sigma 
_2}\delta _{\sigma _3}^{\sigma _4}-\delta _{\sigma _2}^{\sigma _3}\delta 
_{\sigma _1}^{\sigma _4})\delta _{I\cup \sigma ^{\prime }}^{I_1}\delta 
_I^{I_2}\Bigr]  \nonumber \\ 
&&+\delta _{\sigma ^{\prime }}^{\sigma _2}f_{\sigma ^{\prime }}^{\sigma 
_1}f_{\sigma ^{\prime }}^{\sigma _3}\left( f_{\sigma _1}^{\sigma _3}\delta 
_{I\cup (\sigma _1,\sigma _3)}^{I_1}\delta _{I\cup \sigma _4}^{I_2}+\delta 
_{\sigma _3}^{\sigma _4}\delta _{I\cup \sigma _1}^{I_1}\delta 
_I^{I_2}-\delta _{\sigma _1}^{\sigma _4}\delta _{I\cup \sigma 
_3}^{I_1}\delta _I^{I_2}\right) +  \nonumber \\ 
&&-\delta _{\sigma ^{\prime }}^{\sigma _4}f_{\sigma ^{\prime }}^{\sigma 
_1}f_{\sigma ^{\prime }}^{\sigma _3}\left( f_{\sigma _1}^{\sigma _3}\delta 
_{I\cup (\sigma _1,\sigma _3)}^{I_1}\delta _{I\cup \sigma _2}^{I_2}+\delta 
_{\sigma _3}^{\sigma _2}\delta _{I\cup \sigma _1}^{I_1}\delta 
_I^{I_2}-\delta _{\sigma _1}^{\sigma _2}\delta _{I\cup \sigma 
_3}^{I_1}\delta _I^{I_2}\right) \Bigr]\;.  \nonumber 
\end{eqnarray} 
 
Using (\ref{5.3}) we may now define the following vertex-functions for a 
single creation-operator as it occurs in (\ref{5.1a})  
\end{mathletters} 
\begin{mathletters} 
\label{5.7} 
\begin{eqnarray} 
Q_{\frak{D}}^{+}(\sigma ,\sigma ^{\prime }) &=&\sum_{\Gamma ,\Gamma ^{\prime 
}}\lambda _\Gamma \lambda _{\Gamma ^{\prime }}\sum_{I(\sigma \notin 
I)}f_\sigma ^IT_{\Gamma ,I}^{+}T_{I\cup \sigma ,\Gamma ^{\prime 
}}^{}\sum_{I_1,I_2}T_{I_1,\Gamma }^{}T_{\Gamma ^{\prime },I_2}^{+}{\frak{H}}%
_{I_1,I_2}^{\frak{D}}(\sigma ^{\prime })\;,  \label{5.7b} \\ 
\tilde{Q}_{\frak{DD}^{\prime }}^{+}(\sigma ,\sigma ^{\prime }) 
&=&\sum_{\Gamma ,\Gamma ^{\prime }}\lambda _\Gamma \lambda _{\Gamma ^{\prime 
}}\sum_{I(\sigma \notin I)}f_\sigma ^IT_{\Gamma ,I}^{+}T_{I\cup \sigma 
,\Gamma ^{\prime }}^{}\sum_{I_1,I_2}T_{I_1,\Gamma }^{}T_{\Gamma ^{\prime 
},I_2}^{+}{\frak{\tilde{H}}}_{I_1,I_2}^{\frak{DD}^{\prime }}(\sigma ^{\prime 
})  \label{5.7d} 
\end{eqnarray} 
The respective vertices $Q_{\frak{D}}(\sigma ,\sigma ^{\prime })$, $\tilde{Q}%
_{\frak{DD}^{\prime }}(\sigma ,\sigma ^{\prime })$ for 
annihilation-operators are given as  
\end{mathletters} 
\begin{mathletters} 
\label{5.8} 
\begin{eqnarray} 
Q_{\left( \sigma _1\sigma _2\right) }(\sigma ,\sigma ^{\prime }) &=&\left[ 
Q_{\left( \sigma _2\sigma _1\right) }^{+}(\sigma ,\sigma ^{\prime })\right] 
^{*}\;,  \label{5.8a} \\ 
\tilde{Q}_{\left( \sigma _1\sigma _2\right) (\sigma _3\sigma _4)}(\sigma 
,\sigma ^{\prime }) &=&\left[ \tilde{Q}_{\left( \sigma _2\sigma _1\right) 
(\sigma _4\sigma _3)}^{+}(\sigma ,\sigma ^{\prime })\right] ^{*}\;. 
\end{eqnarray} 
 
Now we can determine the connected diagrams in (\ref{5.1a}). First, we have 
the following terms for $i\neq j$%
\end{mathletters} 
\begin{mathletters} 
\label{5.8bb} 
\begin{eqnarray} 
\varepsilon ^2(\vec{q}\,) &=&\frac{N_{\text{G}}}{N_{\vec{q}}}\sum_{\sigma 
_1,\sigma _2}\sqrt{q_{\sigma _1}q_{\sigma _2}}\sum_{\frak{D}_1,\frak{D}%
_2}\left( \tilde{U}_{\frak{D}_1,\frak{D}_2}^{\sigma _1,\sigma _2}(\vec{q}\,)+%
\tilde{U}_{\frak{D}_2,\frak{D}_1}^{\sigma _1,\sigma _2}(\vec{q}\,)\right)  
\label{5.9a} \\ 
&&\times \left[ \left( \hat{1}+\tilde{x}\cdot \tilde{\Omega}(\vec{q}%
\,)\right) \cdot \vec{S}^{+}\right] _{\frak{D}_1}\left[ \left( \hat{1}+%
\tilde{x}\cdot \tilde{\Omega}(\vec{q}\,)\right) \cdot \vec{S}^{-}\right] _{%
\frak{D}_2}\;,  \nonumber \\ 
\varepsilon ^3(\vec{q}\,) &=&\frac{N_{\text{G}}}{N_{\vec{q}}}\sum_{\sigma 
_1,\sigma _2}\sqrt{q_{\sigma _1}q_{\sigma _2}}\sum_{\frak{D}_1,\frak{D}_2,%
\frak{D}_3,\frak{D}_4}V_{\left( \sigma _1\sigma _2\right) \left( \sigma 
_1\sigma _2\right) }^{\frak{D}_1}(\vec{0}\,)\left[ \tilde{x}\cdot \left(  
\hat{1}+\tilde{\Omega}(\vec{0}\,)\cdot \tilde{x}\right) \right] _{\frak{D}_1,%
\frak{D}_2}  \nonumber \\ 
&&\times \left( \tilde{W}_{\frak{D}_3,\frak{D}_4}^{\frak{D}_2}(\vec{q}\,)+%
\tilde{W}_{\frak{D}_4,\frak{D}_3}^{\frak{D}_2}(\vec{q}\,)\right)   \nonumber 
\\ 
&&\times \left[ \left( \hat{1}+\tilde{x}\cdot \tilde{\Omega}(\vec{q}%
\,)\right) \cdot \vec{S}^{+}\right] _{\frak{D}_3}\left[ \left( \hat{1}+%
\tilde{x}\cdot \tilde{\Omega}(\vec{q}\,)\right) \cdot \vec{S}^{-}\right] _{%
\frak{D}_4}\;, \\ 
\varepsilon ^4(\vec{q}\,) &=&\frac{N_{\text{G}}}{N_{\vec{q}}}\sum_{\sigma 
_1,\sigma _2}\sqrt{q_{\sigma _1}q_{\sigma _2}}\sum_{\frak{D}_1,\frak{D}_2,%
\frak{D}_3,\frak{D}_4}V_{\left( \sigma _1\sigma _2\right) \left( \sigma 
_1\sigma _2\right) }^{\frak{D}_1}(\vec{0}\,)\left[ \hat{1}+\tilde{x}\cdot  
\tilde{\Omega}(\vec{0}\,)\right] _{\frak{D}_1,\frak{D}_2}  \nonumber \\ 
&&\times \tilde{\xi}_{\frak{D}_2}^{\frak{D}_3,\frak{D}_4}\left[ \tilde{\Omega%
}(\vec{q}\,)\cdot \vec{S}^{+}\right] _{\frak{D}_3}\left[ \tilde{\Omega}(\vec{%
q}\,)\cdot \vec{S}^{-}\right] _{\frak{D}_4}\;, \\ 
\varepsilon ^5(\vec{q}\,) &=&\frac{N_{\text{G}}}{N_{\vec{q}}}\sum_{\sigma 
_1,\sigma _2,\sigma _1^{\prime },\sigma _2^{\prime }}\sum_{\frak{D}_1,\frak{D%
}_2}\left( Q_{\frak{\bar{D}}_1}^{+}(\sigma _1,\sigma _1^{\prime })Q_{\frak{D}%
_2}(\sigma _2,\sigma _2^{\prime })+Q_{\frak{\bar{D}}_2}^{+}(\sigma _1,\sigma 
_1^{\prime })Q_{\frak{D}_1}(\sigma _2,\sigma _2^{\prime })\right)   \nonumber 
\\ 
&&\times E_{\sigma _1^{\prime },\sigma _2^{\prime }}^{\sigma _1,\sigma _2}(%
\vec{q}\,)\left[ \tilde{\Omega}(\vec{q}\,)\cdot \vec{S}^{+}\right] _{\frak{D}%
_1}\left[ \tilde{\Omega}(\vec{q}\,)\cdot \vec{S}^{-}\right] _{\frak{D}_2}\;, 
\\ 
\varepsilon ^6(\vec{q}\,) &=&\frac{N_{\text{G}}}{N_{\vec{q}}}\sum_{\sigma 
_1,\sigma _2,\sigma _1^{\prime }}\sqrt{q_{\sigma _2}}\sum_{\frak{D}_1,\frak{D%
}_2}\left( Q_{\frak{\bar{D}}_1}^{+}(\sigma _1,\sigma _1^{\prime })\bar{V}%
_{\left( \sigma _1^{\prime }\sigma _2\right) \left( \sigma _1\sigma 
_2\right) }^{\frak{D}_2}(\vec{q}\,)+Q_{\frak{\bar{D}}_1}(\sigma _1,\sigma 
_1^{\prime })V_{\left( \sigma _2\sigma _1^{\prime }\right) \left( \sigma 
_2\sigma _1\right) }^{\frak{D}_2}(\vec{q}\,)\right)   \nonumber \\ 
&&\times \left\{ \left[ \left( \hat{1}+\tilde{x}\cdot \tilde{\Omega}(\vec{q}%
\,)\right) \cdot \vec{S}^{+}\right] _{\frak{D}_2}\left[ \tilde{\Omega}(\vec{q%
}\,)\cdot \vec{S}^{-}\right] _{\frak{D}_1}\right.   \nonumber \\ 
&&+\left. \left[ \tilde{\Omega}(\vec{q}\,)\cdot \vec{S}^{+}\right] _{\frak{D}%
_1}\left[ \left( \hat{1}+\tilde{x}\cdot \tilde{\Omega}(\vec{q}\,)\right) 
\cdot \vec{S}^{-}\right] _{\frak{D}_2}\right\} \;, \\ 
\varepsilon ^7(\vec{q}\,) &=&\frac{N_{\text{G}}}{N_{\vec{q}}}\sum_{\sigma 
_1,\sigma _2,\sigma _1^{\prime }}\sqrt{q_{\sigma _2}}\sum_{\frak{D}_1,\frak{D%
}_2,\frak{D}_3,\frak{D}_4}\left( Q_{\frak{\bar{D}}_1}^{+}(\sigma _1,\sigma 
_1^{\prime })E_{\left( \sigma _1^{\prime }\sigma _2\right) }^{\left( \sigma 
_1\sigma _2\right) }(\vec{0}\,)+Q_{\frak{\bar{D}}_1}(\sigma _1,\sigma 
_1^{\prime })E_{\left( \sigma _2\sigma _1^{\prime }\right) }^{\left( \sigma 
_2\sigma _1\right) }(\vec{0}\,)\right)   \nonumber \\ 
&&\times \left[ \hat{1}+\tilde{\Omega}(\vec{0}\,)\cdot \tilde{x}\right] _{%
\frak{D}_1,\frak{D}_2}\left( W_{\frak{D}_3,\frak{D}_4}^{\frak{D}_2}(\vec{q}%
\,)+W_{\frak{D}_4,\frak{D}_3}^{\frak{D}_2}(\vec{q}\,)\right)   \nonumber \\ 
&&\times \left[ \left( \hat{1}+\tilde{x}\cdot \tilde{\Omega}(\vec{q}%
\,)\right) \cdot \vec{S}^{+}\right] _{\frak{D}_3}\left[ \left( \hat{1}+%
\tilde{x}\cdot \tilde{\Omega}(\vec{q}\,)\right) \cdot \vec{S}^{-}\right] _{%
\frak{D}_4}\;, \\ 
\varepsilon ^8(\vec{q}\,) &=&\frac{N_{\text{G}}}{N_{\vec{q}}}\sum_{\sigma 
_1,\sigma _2,\sigma _1^{\prime }}\sqrt{q_{\sigma _2}}\sum_{\frak{D}_1,\frak{D%
}_2,\frak{D}_3,\frak{D}_4}\left( Q_{\frak{\bar{D}}_1}^{+}(\sigma _1,\sigma 
_1^{\prime })E_{\left( \sigma _1^{\prime }\sigma _2\right) }^{\left( \sigma 
_1\sigma _2\right) }(\vec{0}\,)+Q_{\frak{\bar{D}}_1}(\sigma _1,\sigma 
_1^{\prime })E_{\left( \sigma _2\sigma _1^{\prime }\right) }^{\left( \sigma 
_2\sigma _1\right) }(\vec{0}\,)\right)   \nonumber \\ 
&&\times \tilde{\Omega}_{\frak{\bar{D}}_1}^{\frak{D}_2}(\vec{q}\,)\tilde{\xi}%
_{\frak{D}_2}^{\frak{D}_3,\frak{D}_4}\left[ \tilde{\Omega}(\vec{q}\,)\cdot  
\vec{S}^{+}\right] _{\frak{D}_3}\left[ \tilde{\Omega}(\vec{q}\,)\cdot \vec{S}%
^{-}\right] _{\frak{D}_4}\;, \\ 
\varepsilon ^9(\vec{q}\,) &=&\frac{N_{\text{G}}}{N_{\vec{q}}}\sum_{\sigma 
_1,\sigma _2,\sigma _1^{\prime }}\sqrt{q_{\sigma _2}}\sum_{\frak{D}_1,\frak{D%
}_2}\left( \tilde{Q}_{\frak{\bar{D}}_1\frak{D}_2}^{+}(\sigma _1,\sigma 
_1^{\prime })E_{\left( \sigma _1^{\prime }\sigma _2\right) }^{\left( \sigma 
_1\sigma _2\right) }(\vec{0}\,)+\tilde{Q}_{\frak{\bar{D}}_1\frak{D}%
_2}(\sigma _1,\sigma _1^{\prime })E_{\left( \sigma _2\sigma _1^{\prime 
}\right) }^{\left( \sigma _2\sigma _1\right) }(\vec{0}\,)\right)   \nonumber 
\\ 
&&\times \left[ \tilde{\Omega}(\vec{q}\,)\cdot \vec{S}^{+}\right] _{\frak{D}%
_1}\left[ \tilde{\Omega}(\vec{q}\,)\cdot \vec{S}^{-}\right] _{\frak{D}_2}\;, 
\\ 
\varepsilon ^{10}(\vec{q}\,) &=&\frac{N_{\text{G}}}{N_{\vec{q}}}\sum_{\sigma 
_1,\sigma _2,\sigma _1^{\prime }}\sqrt{q_{\sigma _2}}\sum_{\frak{D}_1}\left( 
Q_{\frak{\bar{D}}_1}^{+}(\sigma _1,\sigma _1^{\prime })E_{\left( \sigma 
_1^{\prime }\sigma _2\right) }^{\left( \sigma _1\sigma _2\right) }(\vec{0}%
\,)+Q_{\frak{\bar{D}}_1}(\sigma _1,\sigma _1^{\prime })E_{\left( \sigma 
_2\sigma _1^{\prime }\right) }^{\left( \sigma _2\sigma _1\right) }(\vec{0}%
\,)\right)   \nonumber \\ 
&&\times \left[ \tilde{\Omega}(\vec{q}\,)\cdot \left( \stackrel{+}{T}\cdot  
\tilde{\Omega}(\vec{q}\,)\cdot \vec{S}^{-}+\stackrel{-}{T}\cdot \tilde{\Omega%
}(\vec{q}\,)\cdot \vec{S}^{+}\right) \right] _{\frak{\bar{D}}_1}\;, \\ 
\varepsilon ^{11}(\vec{q}\,) &=&\frac{N_{\text{G}}}{N_{\vec{q}}}\sum_{\sigma 
_1,\sigma _2}\sqrt{q_{\sigma _1}q_{\sigma _2}}\sum_{\frak{D}_1}V_{\left( 
\sigma _1\sigma _2\right) \left( \sigma _1\sigma _2\right) }^{\frak{D}_1}(%
\vec{0}\,)  \nonumber \\ 
&&\times \left[ \left( \hat{1}+\tilde{x}\cdot \tilde{\Omega}(\vec{q}%
\,)\right) \cdot \left( \stackrel{+}{T}\cdot \tilde{\Omega}(\vec{q}\,)\cdot  
\vec{S}^{-}+\stackrel{-}{T}\cdot \tilde{\Omega}(\vec{q}\,)\cdot \vec{S}%
^{+}\right) \right] _{\frak{D}_1}\;. 
\end{eqnarray} 
Expressions for the diagrams $E_{\frak{D}_2}^{\frak{D}_1}$, $U_{\frak{D}_2%
\frak{D}_3}^{\frak{D}_1}$, $V_{\frak{D}_2\frak{D}_3}^{\frak{D}_1}$, and $%
\bar{V}_{\frak{D}_2\frak{D}_3}^{\frak{D}_1}$ can be found in Appendix (\ref 
{integr}). Using eq. (\ref{4.30}) we derive the connected diagrams in (\ref 
{5.1a}) with $i=j$ as  
\end{mathletters} 
\begin{mathletters} 
\label{5.8ee} 
\begin{eqnarray} 
\varepsilon ^{12}(\vec{q}\,) &=&\frac{N_{\text{G}}}{N_{\vec{q}}}\sum_{\sigma 
_1,\sigma _2}\sqrt{q_{\sigma _1}q_{\sigma _2}}\sum_{\frak{D}_1}\sum_{\Gamma 
^{\prime }}\lambda _{\Gamma ^{\prime }}^2S_{-}(\Gamma ^{\prime 
})V_{\left( \sigma _1\sigma _2\right) \left( \sigma _1\sigma _2\right) }^{%
\frak{D}_1}(\vec{0}\,)\left[ \left( \hat{1}+\tilde{x}\cdot \tilde{\Omega}(%
\vec{0}\,)\right) \cdot \vec{M}(\Gamma ^{\prime })\right] \;,  \label{5.10} 
\\ 
\varepsilon ^{13}(\vec{q}\,) &=&\frac{N_{\text{G}}}{N_{\vec{q}}}\sum_{\sigma 
_1,\sigma _2,\sigma _1^{\prime }}\sqrt{q_{\sigma _2}}\sum_{\frak{D}%
_1}\sum_{\Gamma ^{\prime }}\lambda _{\Gamma ^{\prime }}^2S_{-}(\Gamma 
^{\prime })\left( Q_{\frak{\bar{D}}_1}^{+}(\sigma _1,\sigma _1^{\prime 
})E_{\left( \sigma _1^{\prime }\sigma _2\right) }^{\left( \sigma _1\sigma 
_2\right) }(\vec{0}\,)+Q_{\frak{\bar{D}}_1}(\sigma _1,\sigma _1^{\prime 
})E_{\left( \sigma _2\sigma _1^{\prime }\right) }^{\left( \sigma _2\sigma 
_1\right) }(\vec{0}\,)\right)   \nonumber \\ 
&&\times \left[ \tilde{\Omega}(\vec{0}\,)\cdot \vec{M}(\Gamma ^{\prime 
})\right] _{\frak{\bar{D}}_1}\;. 
\end{eqnarray} 
 
For the evaluation of (\ref{5.1b}) we need the vertex-function for the 
operators  
\end{mathletters} 
\begin{mathletters} 
\label{5.8pp} 
\begin{eqnarray} 
\hat{P}_{\text{G}}\hat{c}_\sigma \hat{S}^{-}\hat{P}_{\text{G}} 
&=&\sum_{\Gamma ,\Gamma ^{\prime }}\lambda _\Gamma \lambda _{\Gamma ^{\prime 
}}\sqrt{S_{-}(\Gamma ^{\prime })}\sum_{I(\sigma \notin I)}f_\sigma 
^IT_{\Gamma ,I}^{+}T_{I\cup \sigma ,\Gamma _{-}^{\prime 
}}^{}\sum_{I_1,I_2}T_{I_1,\Gamma }^{}T_{\Gamma ^{\prime },I_2}^{+}\hat{m}%
_{I_1,I_2}\;,  \label{5.11} \\ 
\hat{P}_{\text{G}}\hat{c}_\sigma ^{+}\hat{S}^{-}\hat{P}_{\text{G}} 
&=&\sum_{\Gamma ,\Gamma ^{\prime }}\lambda _\Gamma \lambda _{\Gamma ^{\prime 
}}\sqrt{S_{-}(\Gamma ^{\prime })}\sum_{I(\sigma \notin I)}f_\sigma 
^IT_{\Gamma ,I\cup \sigma }^{+}T_{I,\Gamma _{-}^{\prime 
}}^{}\sum_{I_1,I_2}T_{I_1,\Gamma }^{}T_{\Gamma ^{\prime },I_2}^{+}\hat{m}%
_{I_1,I_2}\; 
\end{eqnarray} 
with one (or two) incoming or outgoing lines. These functions follow 
directly from eqs. (\ref{5.5}) and (\ref{5.6})  
\end{mathletters} 
\begin{mathletters} 
\label{5.8ff} 
\begin{eqnarray} 
r(\sigma ,\sigma ^{\prime }) &=&\sum_{\Gamma ,\Gamma ^{\prime }}\lambda 
_\Gamma \lambda _{\Gamma ^{\prime }}\sqrt{S_{-}(\Gamma ^{\prime })}%
\sum_{I(\sigma \notin I)}f_\sigma ^IT_{\Gamma ,I}^{+}T_{I\cup \sigma ,\Gamma 
_{-}^{\prime }}^{}\sum_{I_1,I_2}T_{I_1,\Gamma }^{}T_{\Gamma ^{\prime 
},I_2}^{+}{\frak{h}}_{I_2,I_1}(\sigma ^{\prime })\;,  \label{5.12} \\ 
r^{+}(\sigma ,\sigma ^{\prime }) &=&\sum_{\Gamma ,\Gamma ^{\prime }}\lambda 
_\Gamma \lambda _{\Gamma ^{\prime }}\sqrt{S_{-}(\Gamma ^{\prime })}%
\sum_{I(\sigma \notin I)}f_\sigma ^IT_{\Gamma ,I\cup \sigma }^{+}T_{I,\Gamma 
_{-}^{\prime }}^{}\sum_{I_1,I_2}T_{I_1,\Gamma }^{}T_{\Gamma ^{\prime 
},I_2}^{+}{\frak{h}}_{I_1,I_2}(\sigma ^{\prime })\;, \\ 
R_{\sigma _1,\sigma _2}(\sigma ,\sigma ^{\prime }) &=&\sum_{\Gamma ,\Gamma 
^{\prime }}\lambda _\Gamma \lambda _{\Gamma ^{\prime }}\sqrt{S_{-}(\Gamma 
^{\prime })}\sum_{I(\sigma \notin I)}f_\sigma ^IT_{\Gamma ,I}^{+}T_{I\cup 
\sigma ,\Gamma _{-}^{\prime }}^{}\sum_{I_1,I_2}T_{I_1,\Gamma }^{}T_{\Gamma 
^{\prime },I_2}^{+}{\frak{H}}_{I_2,I_1}^{\left( \sigma _2\sigma _1\right) 
}(\sigma ^{\prime })\;, \\ 
R_{\left( \sigma _1\sigma _2\right) }^{+}(\sigma ,\sigma ^{\prime }) 
&=&\sum_{\Gamma ,\Gamma ^{\prime }}\lambda _\Gamma \lambda _{\Gamma ^{\prime 
}}\sqrt{S_{-}(\Gamma ^{\prime })}\sum_{I(\sigma \notin I)}f_\sigma 
^IT_{\Gamma ,I\cup \sigma }^{+}T_{I,\Gamma _{-}^{\prime 
}}^{}\sum_{I_1,I_2}T_{I_1,\Gamma }^{}T_{\Gamma ^{\prime },I_2}^{+}{\frak{H}}%
_{I_1,I_2}^{\left( \sigma _1\sigma _2\right) }(\sigma ^{\prime })\;. 
\end{eqnarray} 
Note that for the vertices of the remaining operators $\hat{P}_{\text{G}}%
\hat{S}^{+}\hat{c}_\sigma \hat{P}_{\text{G}}$ and $\hat{P}_{\text{G}}\hat{S}%
^{+}\hat{c}_\sigma ^{+}\hat{P}_{\text{G}}$ the rules (\ref{5.8}) apply. Now 
we are able to summarize the contributions (\ref{5.1b}) as  
\end{mathletters} 
\begin{mathletters} 
\label{5.8gg} 
\begin{eqnarray} 
\varepsilon ^{14}(\vec{q}\,) &=&\frac{N_{\text{G}}}{N_{\vec{q}}}\sum_{\sigma 
_1,\sigma _2,\sigma _1^{\prime }}\sqrt{q_{\sigma _2}}\sum_{\frak{D}_1}\left( 
r^{+}(\sigma _1,\sigma _1^{\prime })\bar{V}_{\left( \sigma _1^{\prime 
}\sigma _2\right) \left( \sigma _1\sigma _2\right) }^{\frak{D}_1}(\vec{q}%
\,)+r(\sigma _1,\sigma _1^{\prime })V_{\left( \sigma _2\sigma _1^{\prime 
}\right) \left( \sigma _2\sigma _1\right) }^{\frak{D}_1}(\vec{q}\,)\right)   
\nonumber \\ 
&&\times \left[ \left( \hat{1}+\tilde{x}\cdot \tilde{\Omega}(\vec{q}%
\,)\right) \cdot \vec{S}^{-}\right] _{\frak{D}_1}+c.c.\;,  \label{5.14} \\ 
\varepsilon ^{15}(\vec{q}\,) &=&\frac{N_{\text{G}}}{N_{\vec{q}}}\sum_{\sigma 
_1,\sigma _2,\sigma _1^{\prime }}\sqrt{q_{\sigma _2}}\sum_{\frak{D}_1}\left( 
R_{\frak{\bar{D}}_1}^{+}(\sigma _1,\sigma _1^{\prime })E_{\left( \sigma 
_1^{\prime }\sigma _2\right) }^{\left( \sigma _1\sigma _2\right) }(\vec{0}%
\,)+R_{\frak{\bar{D}}_1}(\sigma _1,\sigma _1^{\prime })E_{\left( \sigma 
_2\sigma _1^{\prime }\right) }^{\left( \sigma _2\sigma _1\right) }(\vec{0}%
\,)\right)   \nonumber \\ 
&&\times \left[ \tilde{\Omega}(\vec{q}\,)\cdot \vec{S}^{-}\right] _{\frak{D}%
_1}+c.c\;, \\ 
\varepsilon ^{16}(\vec{q}\,) &=&\frac{N_{\text{G}}}{N_{\vec{q}}}\sum_{\sigma 
_1,\sigma _2,\sigma _1^{\prime },\sigma _2^{\prime }}\sum_{\frak{D}_1}\left( 
r^{+}(\sigma _1,\sigma _1^{\prime })Q_{\frak{\bar{D}}_1}(\sigma _2,\sigma 
_2^{\prime })+r(\sigma _2,\sigma _2^{\prime })Q_{\frak{\bar{D}}%
_1}^{+}(\sigma _1,\sigma _1^{\prime })\right) E_{\left( \sigma _1^{\prime 
}\sigma _2^{\prime }\right) }^{\left( \sigma _1\sigma _2\right) }(\vec{q}\,) 
\nonumber \\ 
&&\times \left[ \tilde{\Omega}(\vec{q}\,)\cdot \vec{S}^{-}\right] _{\frak{D}%
_1}+c.c.\;. 
\end{eqnarray} 
The contributions from eq. (\ref{5.1c}) are  
\end{mathletters} 
\begin{equation} 
\varepsilon ^{17}(\vec{q}\,)=\frac{N_{\text{G}}}{N_{\vec{q}}}\sum_{\sigma 
_1,\sigma _2,\sigma _1^{\prime },\sigma _2^{\prime }}\left( r^{+}(\sigma 
_1,\sigma _1^{\prime })\left( r^{+}(\sigma _2,\sigma _2^{\prime })\right) 
^{*}+\left( r(\sigma _1,\sigma _1^{\prime })\right) ^{*}r(\sigma _2,\sigma 
_2^{\prime })\right) E_{\left( \sigma _1^{\prime }\sigma _2^{\prime }\right) 
}^{\left( \sigma _1\sigma _2\right) }(\vec{q}\,)\;.  \label{5.15} 
\end{equation} 
Finally, we need the vertex-function for the operator $\hat{P}_{\text{G}}%
\hat{S}^{+}\hat{c}_\sigma ^{+}\hat{S}^{-}\hat{P}_{\text{G}}$ in (\ref{5.1d}) 
with one outgoing line  
\begin{equation} 
l(\sigma ,\sigma ^{\prime })=\sum_{\Gamma ,\Gamma ^{\prime }}\lambda _\Gamma 
\lambda _{\Gamma ^{\prime }}\sqrt{S_{-}(\Gamma ^{\prime })S_{-}(\Gamma 
)}\sum_{I(\sigma \notin I)}f_\sigma ^IT_{\Gamma _{-},I}^{+}T_{I\cup \sigma 
,\Gamma _{-}^{\prime }}^{}\sum_{I_1,I_2}T_{I_1,\Gamma }^{}T_{\Gamma ^{\prime 
},I_2}^{+}{\frak{h}}_{I_1,I_2}(\sigma ^{\prime })\;. 
\end{equation} 
in (\ref{5.1d}). This gives us the contribution from eq. (\ref{5.1d}),  
\begin{equation} 
\varepsilon ^{18}(\vec{q}\,)=\sum_{\sigma _1,\sigma _2,\sigma _1^{\prime }}%
\sqrt{q_{\sigma _2}}l(\sigma _1,\sigma _1^{\prime })E_{\left( \sigma 
_1^{\prime }\sigma _2\right) }^{\left( \sigma _1\sigma _2\right) }(\vec{0}%
\,)+c.c. 
\end{equation} 
Altogether we may write the expectation value for the kinetic energy (\ref 
{5.1}) as  
\begin{equation} 
\frac{\left\langle \Psi _{\vec{q}}^{\text{G}}\left| \hat{H}_1\right| \Psi _{%
\vec{q}}^{\text{G}}\right\rangle }{N_{\vec{q}}}=E_{\text{kin}%
}+\sum_{c=1}^{18}\varepsilon ^c(\vec{q}\,)\;. 
\end{equation}

\begin{figure}[tbp] 
\caption{Multi-orbital RPA-diagrams for the matrix $\tilde{\Omega}$.} 
\label{fig1} 
\end{figure} 

\begin{figure}[tbp] 
\caption{Phase diagram as a function of $U$ and $J$ for the Hartree-Fock-Stoner theory (HF) and the Gutzwiller wave function (GW) for (a) $n_{\sigma}=0.29$ and (b) $n_{\sigma}=0.35$; PM: paramagnet, FM: ferromagnet.} 
\label{fig2} 
\end{figure} 

\begin{figure}[tbp] 
\caption{Variational spin wave dispersion in (100)-direction, for the two-band model; $n_{\sigma}=0.29$, $J=0.2U$, and the values $U/eV=7.8,\,10,\,12,\,13.6$ correspond to $m=0.12,\,0.20,\,0.26,\,0.28$. The inset shows the Variational spin wave dispersion in (100) and (110)-direction for $m=0.2$ and $m=0.28$} 
\label{fig3} 
\end{figure} 
  
\begin{figure}[tbp] 
\caption{All contributing diagrams in the evaluation of the atomic energy. The diagrams $\tilde{\Phi}(\vec{q})$ and $\tilde{\Psi}(\vec{q})$ are defined as $\tilde{\Phi}=\hat{1}-\tilde{\Omega}\cdot\tilde{x}$, $\tilde{\Psi}=\hat{1}+\tilde{x}\cdot\tilde{\Omega}$. Dotted and solid lines indicate the arguments $\vec{0}$ and $\vec{q}$, respectively.} 
\label{fig4} 
\end{figure} 
 
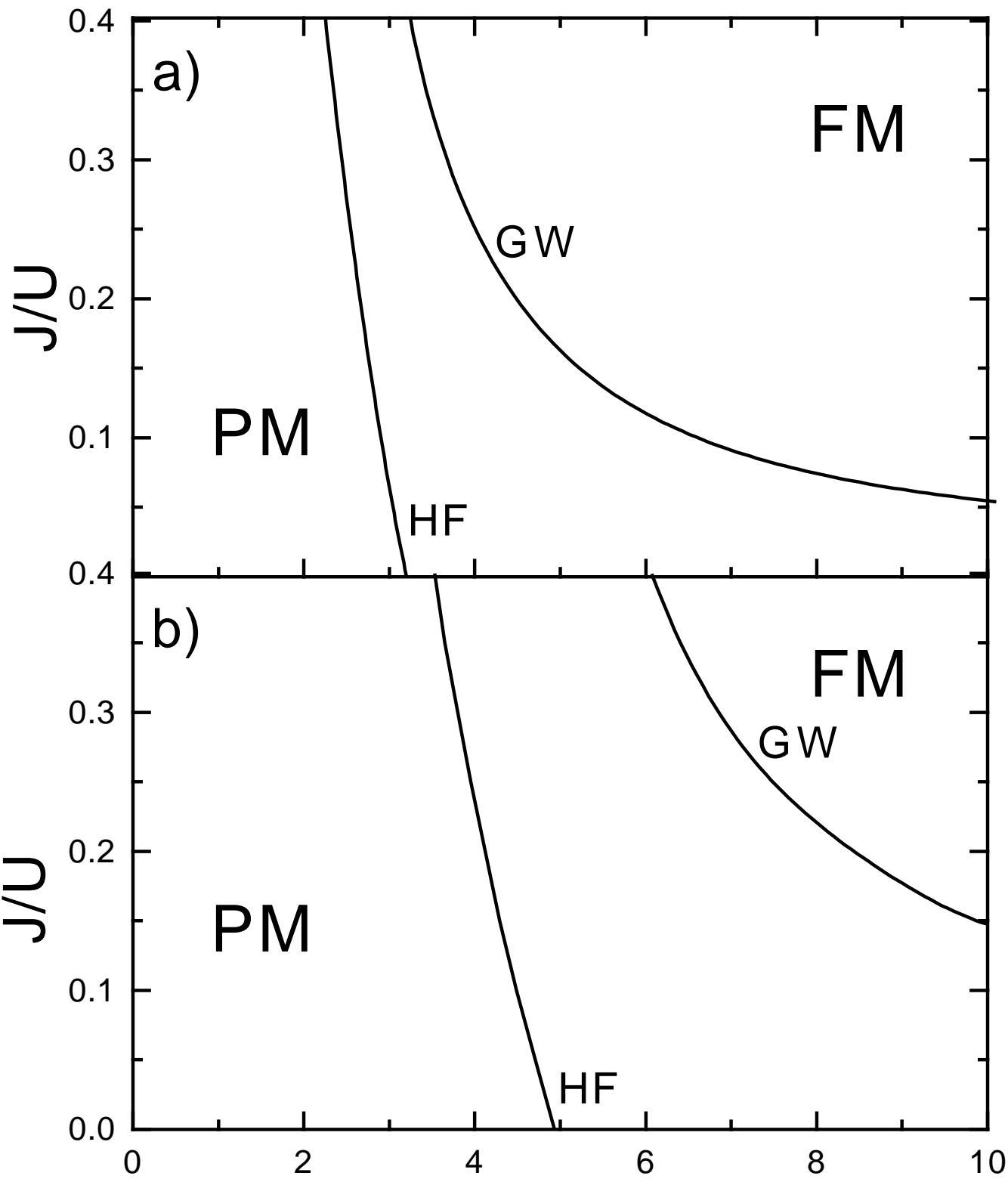
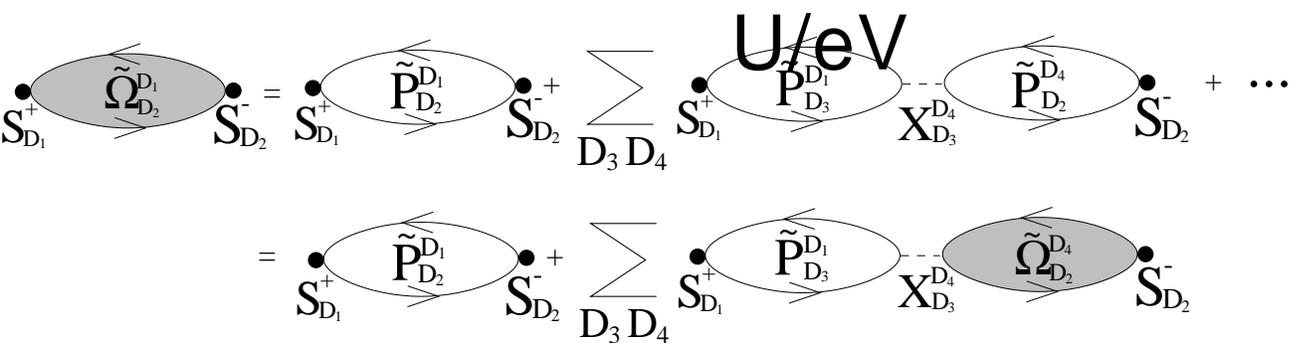
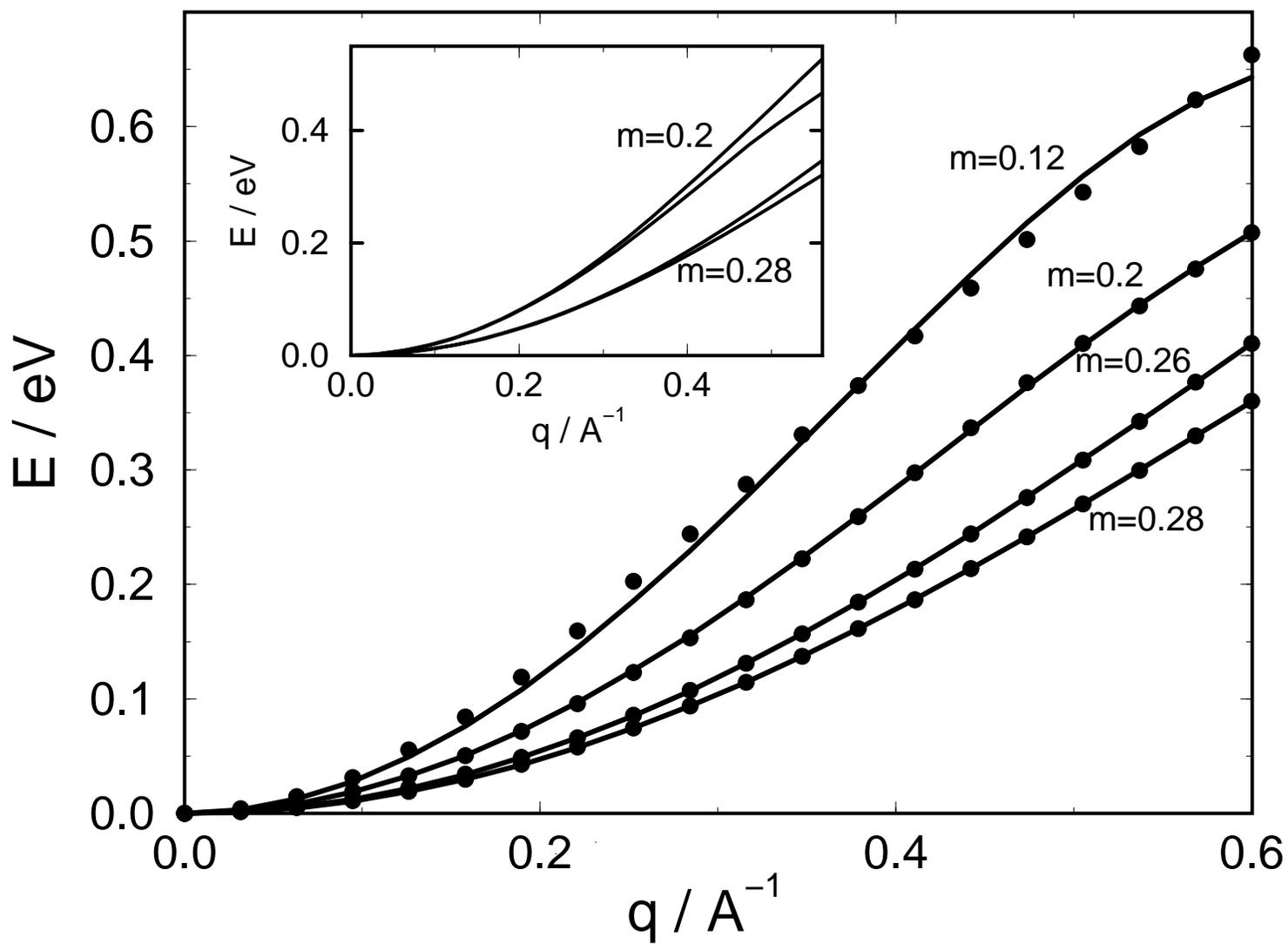
\end{document}